%% file: FrictionFlag.tex
\documentclass[preprint, aps, pre,showpacs]{revtex4-1} 
\usepackage{graphicx, bm, bbm,amsmath, amssymb, epsfig,multirow}

\begin{document}
\input{macros}

\title{Dynamics and locomotion of flexible foils in a frictional environment}
\author{Xiaolin Wang$^{1,*}$,Silas Alben$^{1,}$}
\email{wxiaolin@umich.edu, alben@umich.edu}
\affiliation{$^1$Department of Mathematics, University of Michigan}

\date{\today}

\begin{abstract}
Over the past few decades, oscillating flexible foils have been used to study the physics of organismal propulsion in different fluid environments. Here we extend this work to a study of flexible foils in a frictional environment. When the foil is oscillated by heaving at one end but not allowed to locomote freely, the dynamics change from periodic to non-periodic and chaotic as the heaving amplitude is increased or the bending rigidity is decreased. For friction coefficients lying in a certain range, the transition passes through a sequence of $N$-periodic and asymmetric states before reaching chaotic dynamics. Resonant peaks are damped and shifted by friction and large heaving amplitudes, leading to bistable states.

When the foil is allowed to locomote freely, the horizontal motion smoothes the resonant behaviors. For moderate frictional coefficients, steady but slow locomotion is obtained. For large transverse friction and small tangential friction corresponding to wheeled snake robots, faster locomotion is obtained. Traveling wave motions arise spontaneously, and and move with horizontal speed that scales as transverse friction to the 1/4 power and input power that scales as transverse friction to the 5/12 power. These scalings are consistent with a boundary layer form of the solutions near the foil's leading edge.

\end{abstract}
\keywords{snake locomotion, frictional medium, flexible foil, boundary layer}
\maketitle
\section{Introduction}
Snake locomotion has long been an interesting topic for biologists, engineers and applied mathematicians \cite{gray1946mechanism,Gray, Hirose, Hu09,transeth09,guo2008limbless}, as the lack of limbs distinguishes snake kinematics from other common modes of
locomotion including flying, swimming and walking \cite{dickinson,triantafyllou2000hydrodynamics, wang2005dissecting}, and exhibits unique dynamic behavior \cite{lillywhite2014snakes}. Snakes
gain thrust from the surrounding environment with a variety of gaits, including slithering,
sidewinding, concertina motion, and rectilinear progression \cite{Hu09}. Among these gaits, undulatory motion is one of the most common and is used by many different
limbless animals. Some examples include swimming at low and high Reynolds numbers \cite{purcell1977life,triantafyllou2000hydrodynamics,yu2006experimental,cohen2010swimming,curet2011mechanical,williams2014self} and moving in granular media
\cite{juarez2010motility,peng2016characteristics,peng2017propulsion}. Some limbed animals also use body undulations instead of their limbs in granular media (i.e. the sandfish lizard) \cite{maladen2009undulatory}.

The dynamical features and locomotor behaviors of different undulatory organisms depend on how they interact with the environment, and in particular, depend on the type of
thrust gained from the environment. In fluids, propulsive forces are obtained as a balance of fluid forces and bending rigidity when the body of a flexible swimmer
 is actuated \cite{wiggins1998flexive,lauga2007floppy}. For snakes, previous work showed how self-propulsion arises through effects such as Coulomb frictional forces and internal
viscoelasticity \cite{guo2008limbless,Hu09}.
Here Coulomb friction depends on the direction of the velocity but not its magnitude, and is difficult to analyze theoretically in a fully coupled model, where the shape of the body and its velocity need to be solved simultaneously.
 Previous work approached
this problem by prescribing the motion of the snake in certain functional forms \cite{guo2008limbless, Hu09, jing2013optimization,alben2013optimizing, wang2014optimizing}. In \cite{guo2008limbless}, Guo and Mahadevan studied the effects of internal elasticity, muscular activity and other physical parameters on
locomotion for prescribed sinusoidal motion and sinusoidal and square-wave internal bending
moments. Hu and Shelley \cite{Hu09} assumed a sinusoidal traveling-wave body curvature, and
computed the snake speed and locomotor efficiency with different traveling wave
amplitudes and wavelengths. Their results showed good agreement with biological snakes.
Jing and Alben found time-periodic kinematics of 2- and 3-link bodies that are optimal for efficiency, including traveling wave kinematics in the 3-link case \cite{jing2013optimization}. Alben \cite{alben2013optimizing} found the kinematics of general smooth bodies that are optimal for efficiency for various friction coefficients. These included transverse undulation, ratcheting motions, and direct-wave locomotion. Theoretical analysis showed that with large transverse friction, the optimal motion is a traveling wave (i.e. transverse undulation) with an amplitude that scales as the transverse friction coefficient to the -1/4 power. Wang \etal studied transverse undulation on inclined surfaces with prescribed triangular and sinusoidal deflection waves. They found numerically and theoretically how the optimal wave amplitude depends on the frictional coefficients and incline angles in the large transverse friction coefficient regime \cite{wang2014optimizing}.

The study of flexible foils in a frictional medium applies to various situations where Coulomb friction applies. One example is a granular medium where the resistive force can be modeled using Coulomb friction \cite{maladen2009undulatory,hatton2013geometric}. In the regime of slow movement,
this force model is consistent with experimental results for sand lizards \cite{maladen2009undulatory}. Peng \etal applied this model to investigate the locomotion of a slender swimmer in a granular medium by prescribing a travelling wave body shape, and found the optimal swimming speed and efficiency versus wave number \cite{peng2016characteristics}. In a more recent work, Peng \etal considered propulsion in a granular medium
under the effect of both elasticity and frictional force. They proposed a model where a rigid rod is connected to a torsional spring under a displacement actuation \cite{peng2017propulsion}, studied the effects of actuation amplitude
and spring stiffness on propulsive dynamics, and found the maximum thrust that could be obtained.

In this work, we propose a fully-coupled model to solve for the dynamics and locomotion of an elastic foil by a generalization of previous work \cite{alben2013optimizing, wang2014optimizing}. The snake is described as a 1D flexible
foil whose curvature is a function of arclength and time. Both the internal bending rigidity and frictional forces are included in the model. A sinusoidal heaving motion is prescribed at the leading edge to move the foil, similarly to other recent experiments and numerical studies \cite{yu2006experimental, lauder2011bioinspiration, lauder2011robotic}. The foil moves according to nonlinear force balance equations which will be solved numerically. To simplify the discussion, we first consider the case where the foil is actuated at the leading edge but not free to move (fixed base), and study the dynamics of the foil
at different parameters. Then we allow the foil to move freely in the horizontal direction and consider
the kinematics of the locomotion. This approach has been used to study an actuated elastica in a viscous fluid \cite{lauga2007floppy}, where the fixed base case was used to derive a scaling law for the propulsive forces and the free translation case was used to calculate the swimming speed.
In the fixed base case, we will study some key dynamic phenomena including resonant vibrations at certain bending rigidities, effects of nonlinearity due to large heaving amplitude and geometrical nonlinearities, and transitions from periodic to non-periodic states. Similar phenomena have been discussed previously for mechanical vibrations \cite{landau1986theory} and for swimming in a fluid medium \cite{lauder2012passive,paraz2016thrust, spagnolie2010surprising}, but not
in a frictional environment.
In the freely locomoting case, we will focus on the speed and the input power for the locomotion, and discuss how they scale with physical parameters.

We note that passive flexible foils are a useful model system and have been used in other problems including locomotion in fluids and granular media for a few reasons: they require a very simple control, such as harmonic heaving or pitching at the leading edge, they allow for generic body-environment interactions to arise spontaneously, and they mimic the behavior of flexible bodies and tails commonly used for propulsion in swimming and crawling organisms. Scaling laws for locomotion have been derived for low and high-Reynolds number swimming systems by means of analysis, simulation and experiments \cite{lauga2007floppy, gazzola2014scaling,alben2012dynamics,quinn2014scaling}. Gazzola \etal \cite{gazzola2014scaling} derived scaling laws that link swimming speed to tail beat
amplitude and frequency and fluid viscosity for inertial aquatic swimmers. Two different scalings were given for laminar flow and high Reynolds number turbulent flow. Alben \etal \cite{alben2012dynamics} numerically and analytically studied the scalings of local maxima in the swimming speed of heaving flexible foils, which indicated that the performance of the
propulsors depends on fluid-structure resonances. This was also studied in an experiment by Quinn \etal \cite{quinn2014scaling} using rectangular panels in a water channel. In this work, we will use passive flexible foils to study similar phenomena in a frictional environment.

This paper is organized as follows: The foil and friction models and the numerical methods are described in Section II. The fixed base cases are discussed in Section III followed by the free locomotion cases in Section IV. Conclusions are given in Section V.
\section{Modelling}
\subsection{Foil and Friction Models}
We consider here the motion of a flexible foil in a frictional environment with a prescribed heaving motion at the leading edge. The foil has chord length $L$,
mass per unit length $\rho$, and bending rigidity $B$. The foil thickness is assumed to be much smaller than its
length and width, and therefore we model it as a 1D inextensible elastic sheet.

The instantaneous position of the foil is described as $\z(s,t)=x(s,t)+iy(s,t)$, where $s$ is arclength. Assuming an Euler-Bernoulli model for the foil,
 the governing equation for $\z$ is:
\bq
\rho\partial_{tt}\z(s,t)=\partial_s(T(s,t)\hat{s})-B\partial_s(\partial_s\kappa(s,t)\hat{n})+f(s,t)
\label{eq:beameq}
\eq
Here $T(s,t)$ is a tension force accounting for the inextensibility, and $\hs=\ds\zeta$ and $\hn=i\hs$ represent the unit vectors tangent
and normal to the foil, respectively. $f(s,t)$ is the frictional force per unit length, and from previous work \cite{Hu09,alben2013optimizing,wang2014optimizing}, it can be described as:
\bq
f(s,t) = -\rho g\mu_t(\hptZ\cdot\hn)\hn-\rho g\mu_f(\hptZ\cdot\hs)\hs.
\eq
 The hats denote normalized vectors and we define $\hptZ$ to be $0$ when the snake velocity is $0$.
The friction coefficients are $\mu_f$ and $\mu_t$ for motions in the tangential $(\pm\hs)$ and transverse $(\pm\hn)$ directions, respectively.

At the leading edge, the transverse position of the foil is prescribed as a sinusoidal function with frequency $\omega$ and amplitude $A$, and the tangent angle $\theta$ is set to zero:
\bq
\zeta(0,t)=X_0(t)+iA\sin(\omega t),\ \theta(0,t)=0
\eq
With a fixed base, $X_0(t)\equiv 0$. For a locomoting body, $X_0(t)$ is computed by assuming no horizontal force is applied at the leading edge: $T(0,t)\equiv 0$. Similar clamped boundary condition have been used in previous experiments and models to study swimming by flexible foils \cite{alben2012dynamics, lauder2011robotic, lauder2012passive,paraz2016thrust}. We note that other choices of boundary conditions have also been applied. For example, a pitching motion where the tangent angle is a sinusoidal function of time while the transverse displacement is fixed to be zero, is another a popular choice \cite{wiggins1998flexive,yu2006experimental,lauga2007floppy}. A torsional flexibility model, in which a torsional spring is connected to a rigid plate at the leading edge, has also been applied in some works \cite{moore2014analytical,peng2017propulsion}.

At the trailing edge, the foil satisfies free-end conditions, which state that the tension force, shearing force and bending moment are all zero:
\bq
T(s=L,t)=\partial_s\kappa(s=L,t)=\kappa(s=L,t)=0 \label{eq:beambd2}
\eq

\subsection{Nondimensionalization}
We nondimensionalize the governing equations and boundary conditions (\ref{eq:beameq}) - (\ref{eq:beambd2})
by the chord length $L$ and the period of the heaving motion, $\tau=\displaystyle{2\pi}{\omega}$.
We obtain the following dimensionless parameters:
 \bq
 \tilde{B}=\frac{B\tau^2}{\rho L^4},\quad\tilde{\mu}_t=\frac{g\mu_t\tau^2}{L},\quad\tilde{\mu}_f=\frac{g\mu_f\tau^2}{L},\quad \tilde{A}=\frac{A}{L}\nonumber
 \eq
and the following dimensionless equations:
\bq
\dtt\zeta(s,t)=\ds(T\hs)-\tilde{B}\ds(\ds\kappa\hn)+f(s,t)\label{eq:nonbeam}
\eq
\bq
f(s,t) = -\tilde{\mu}_t(\hptZ\cdot\hn)\hn-\tmu_f(\hptZ\cdot\hs)\hs
\eq
with the boundary conditions
\bq
\zeta(0,t)=\tilde X_0(t)+i\tilde{A}\sin(2\pi t),\ \theta(0,t)=0;\quad T(1,t)=\partial_s\kappa(1,t)=\kappa(1,t)=0.\label{eq:nonbeambd}
\eq
Now the heaving motion has a period of 1.
For simplicity, we will use the original notation for the parameters instead of the tilded ones in the following sections.

\subsection{Numerical Methods}
We couple equations and boundary conditions (\ref{eq:nonbeam}) - (\ref{eq:nonbeambd}) together to solve for the positions of the body $\zeta(s,t)$ at each time step.
This requires solving a nonlinear system $\mathbf{F}(\mathbf{x})=0$ at each time step and Broyden's method \cite{ralston2012first} is used
 to do so, which essentially requires the evaluation of the function $\mathbf{F}(\mathbf{x})$ for a given $\mathbf{x}$.

At each time step $t_n$, given $\kappa_n$, we first obtain $\zeta_n$:
\bqs
\zeta_n(s,t)=\int_0^se^{i\theta_n}ds',\ \theta_n(s,t)=\int_0^s\kappa_nds'.
\eqs
We then discretize the time derivatives with a second-order (BDF) discretization using previous time step solutions.
If we dot both sides of equation (\ref{eq:nonbeam}) with $\hs$ and integrate from $s=1$, the tension force can be computed as:
\bq
T_n(s,t)=-\frac{1}{2}B\kappa_n^2(s,t)+\int_1^s \dtt\zeta_n\cdot\hs+\mu_f(\hptZ_n\cdot\hs) ds'
\eq
If we dot the same terms with $\hn$ and integrate from $s=1$, we obtain the curvature as:
\bq
\kappa(s,t)=\int_1^s\frac{1}{B}\int_1^{s'}-\dtt\zeta_n\cdot\hn+T_n\kappa_n-\mu_t(\hptZ_n\cdot\hn)ds''ds'
\eq
And the function we drive to zero using Broyden's method is written as:
\bq
\mathbf{F}(\mathbf{x})=\kappa(s,t)-\kappa_n(s,t)
\eq
Most terms in the integrals can be computed using the trapezoidal rule. However, when the velocity of the foil is close to zero, both discretized version of $\displaystyle\int\hptZ\cdot\hs$
and $\displaystyle\int\hptZ\cdot\hn$ can be unbounded locally using a uniform mesh as shown in \cite{alben2013optimizing}, unless the meshes
are locally adaptive. In order to achieve
convergence as well as second-order accuracy of the numerical integration with a uniform mesh,
we use a different approach to evaluate the integral as suggested in \cite{alben2013optimizing}.

We use the notation $u_s=\partial_s\zeta\cdot\hs$, and $u_n=\partial_s\zeta\cdot\hn$ to denote the tangential and normal components of the velocity. Therefore the
normalized velocity components can be rewritten as
\bq
\hptZ\cdot\hs=\frac{u_s}{\sqrt{u_s^2+u_n^2}},\ \hptZ\cdot\hn=\frac{u_n}{\sqrt{u_s^2+u_n^2}}
\eq
On each subinterval $[a,b]$, if we approximate $u_s$ and $u_n$ by linear approximations $As+B$ and $Cs+D$, then the integrals are of the form:
\bq
\int_a^b \frac{As+B}{\sqrt{(As+b)^2+(Cs+D)^2}}
\eq
This integral can then be evaluated analytically and a second-order accuracy is achieved. More details about this integration approach can be found in \cite{alben2013optimizing}.
 Since the discretization requires two previous time step solutions, we need to adjust the method at the first step, which we describe in Appendix A.

\section{Fixed Base}
Because little is known about the dynamical behavior of an elastic body in a frictional medium, we begin with
the simplified case when the base is fixed $(X_0(t) \equiv 0)$, and then study the case of a freely locomoting foil in next section.
\subsection{Zero Friction}
When the heaving amplitude $A$ is small, the nonlinear equation (\ref{eq:nonbeam}) can be linearized by assuming $s\approx x$ and
$\zeta(x,t)\approx x+iy(x,t)$. Therefore, equation (\ref{eq:nonbeam}) becomes:
\bq
\dtt y(x,t)=-B\partial_x^4y \label{eq:linbeam1},
\eq
with friction coefficients set to zero. The boundary conditions become:
\bq
y(0,t)=A\sin(2\pi t),\ \partial_xy(0,t)=0;\quad \partial_x^2y(1,t)=\partial_x^3y(1,t)=0
\eq
and the initial conditions are:
\bq
y(x,0)=y_0(x),\quad \dt y(x,0)=\dt y_0(x)\label{eq:initcond}
\eq
This equation can be solved analytically by using separation of variables as shown in Appendix B. The deflection of the linearized solution is given by
\bq
y(x,t)=\sum\limits_{i=1}^\infty A_i(t)\phi_i(x)+A\sin(2\pi t)
\eq
where $\phi_i(x)$ is an eigenfunction and $A_i(t)$ is a time-dependent coefficient (see details in Appendix B). For arbitrary initial condition, the shape of the foil is non-periodic in time, with a superposition of the heaving frequency 1 and natural frequencies $\sqrt{\lambda_i}$. The linearized model approximates the foil motion well when the heaving amplitude $A$ is small, and we validate our numerical scheme by comparing with the linearized model in Appendix B.

\subsection{Nonzero Friction}

The frictional force adds damping to the system, and therefore damps out the initial transient, leaving a periodic solution (at small enough $A$) with energy input by heaving and removed by friction.
A larger heaving amplitude $A$ enhances other vibration modes which introduces more frequencies into the system and eventually results in a non-periodic motion.
A more flexible foil (i.e., a smaller $B$) also leads to a non-periodic solution. Therefore, the parameters $A$, $B$ and the two friction coefficients will compete
to determine whether the vibration will be periodic or not. For simplicity, we only consider homogeneous friction coefficients $\mu=\mu_t=\mu_f$ in this section.
For snakes and snake-like robots, these parameters are generally not the same \cite{Hu09}.

We first consider the effect of varying $B$ and $\mu$ with fixed heaving amplitude $A$. In figure \ref{fig:periodiag}, we plot a diagram of different dynamical states
after 200 periods with $A=0.3$. \Etwofigs{periodiag}{periodiaglargemu}{\label{fig:periodiag}Diagram of periodicity of the vibration for (a) smaller frictional coefficients $\mu<1$; (b)
larger frictional cofficients with $A=0.3$ and various $B$. $\circ$: non-periodic vibration; $+$: symmetric vibration of period 1; $\triangleright$:
asymmetric vibration of period 1;
$\times$: symmetric vibration of period 3; $\square$: asymmetric vibration of period 5;
$\bigtriangleup$:symmetric vibration of period 6; $\bigtriangledown$: symmetric vibration of period 9; $\triangleleft$: asymmetric vibration of
period 10; $\Diamond$ asymmetric vibration of period 13. }{2.7}{2.7}
When $\mu$ is relatively small ($<0.7$ in this case), the vibration transitions directly from
a periodic (with period 1) to a non-periodic chaotic state
as the foil becomes more flexible as shown in figure \ref{fig:periodiag}(a). The transition value of $B$ is almost invariant for a large range of $\mu$, becoming
slightly smaller as $\mu$ increases. However, when $\mu$ is moderate ($0.7<\mu<7$ for this case), a transition region is observed between the non-periodic and period-1
 states as shown in panels (a) and (b).
The transition region has complex dynamical behaviors with a mixture of different periodic and non-periodic states.

\begin{figure}[!]
          \begin{center}
          \setlength{\tabcolsep}{0in}
          \begin{tabular}{ccccc}
\Dfig{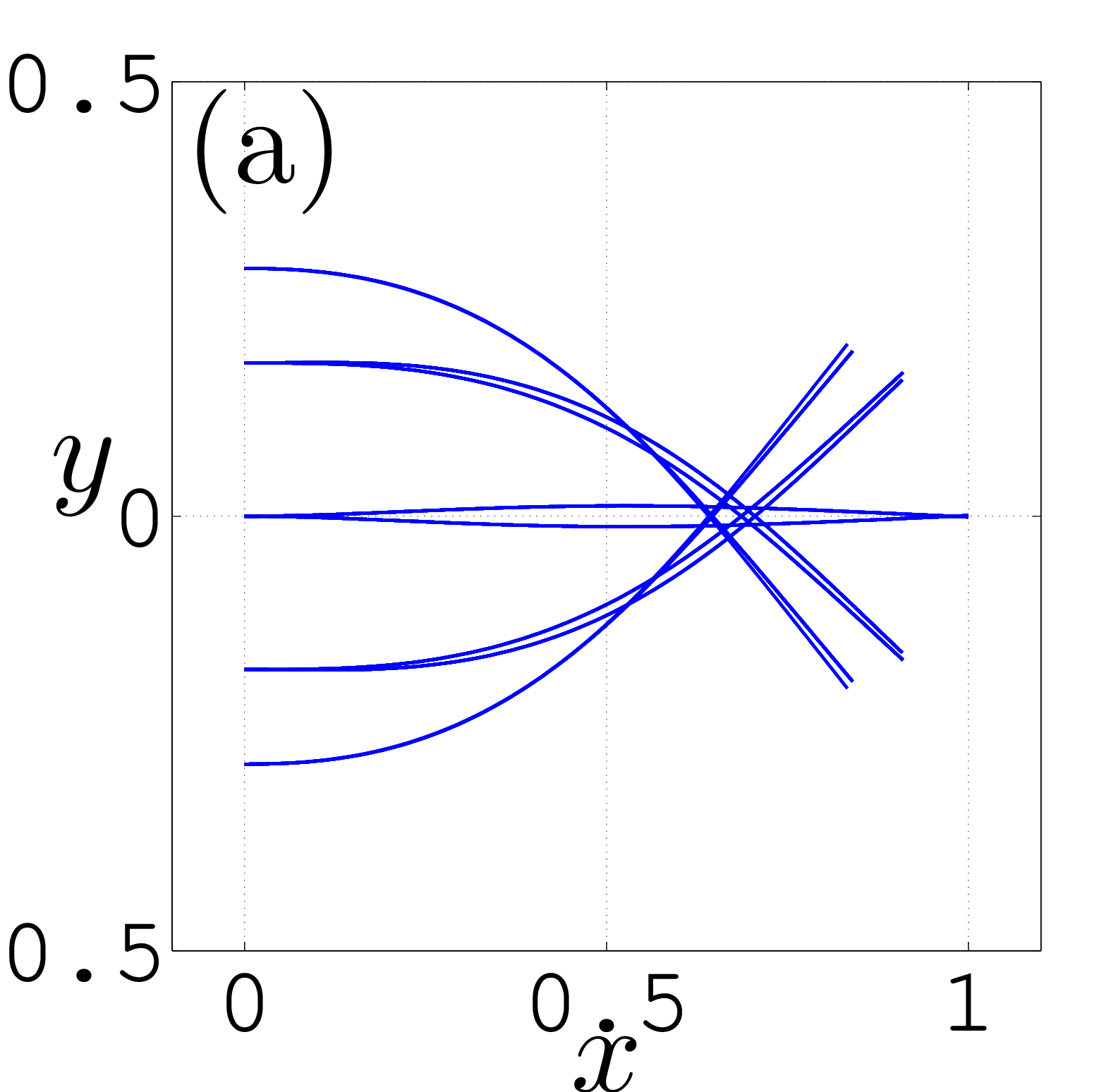}{1.2} & \Dfig{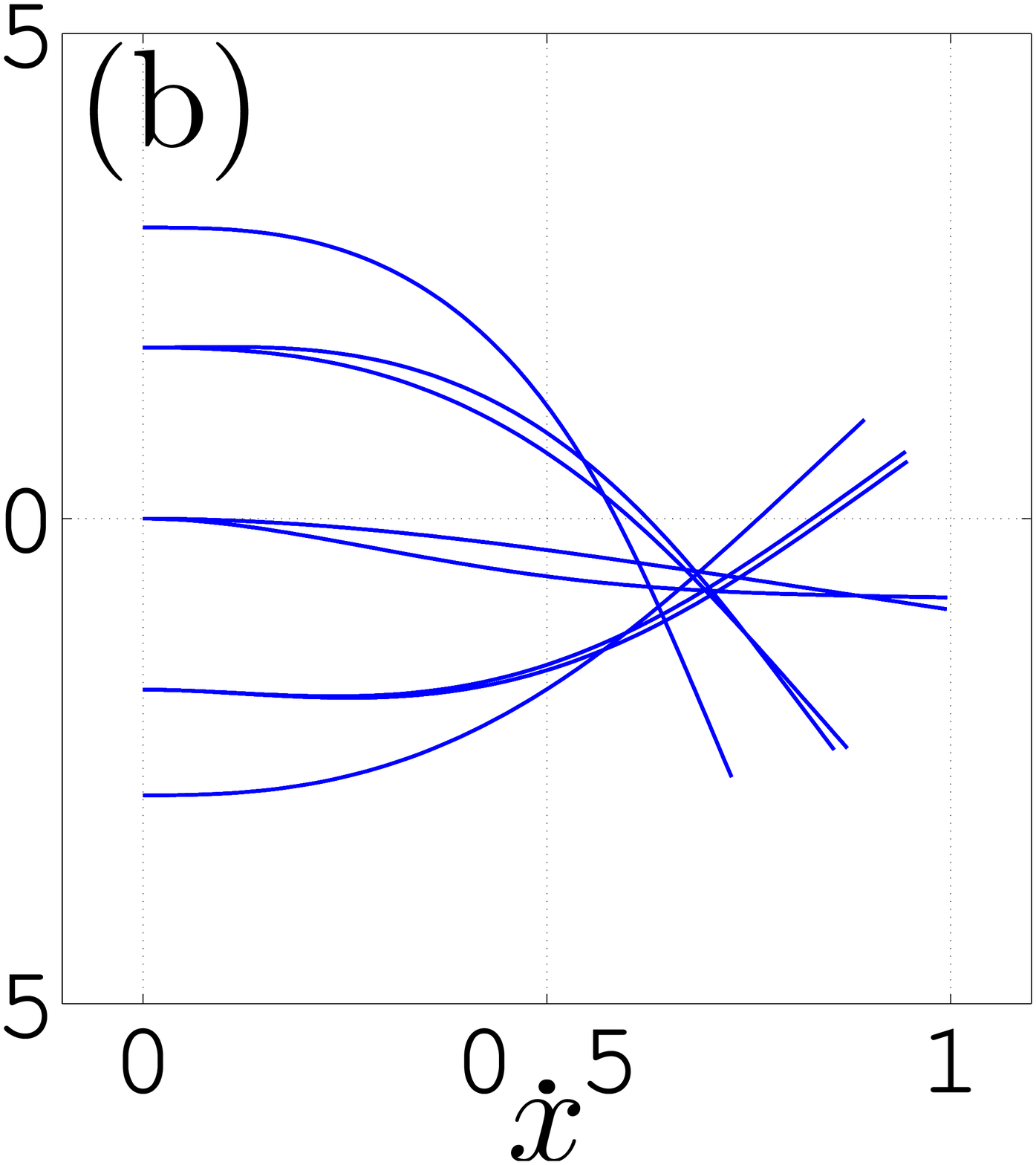}{1.2}& \Dfig{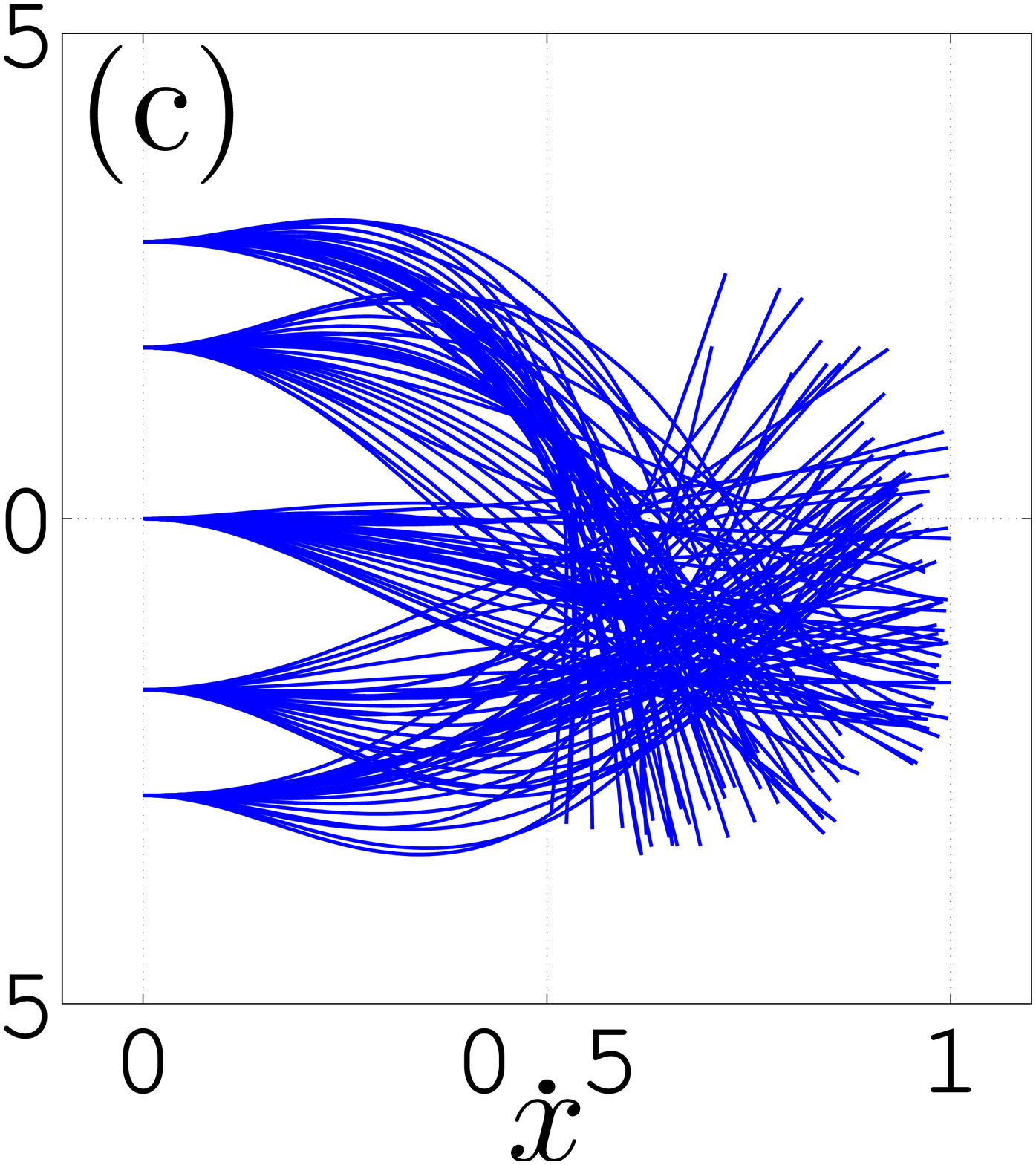}{1.2} & \Dfig{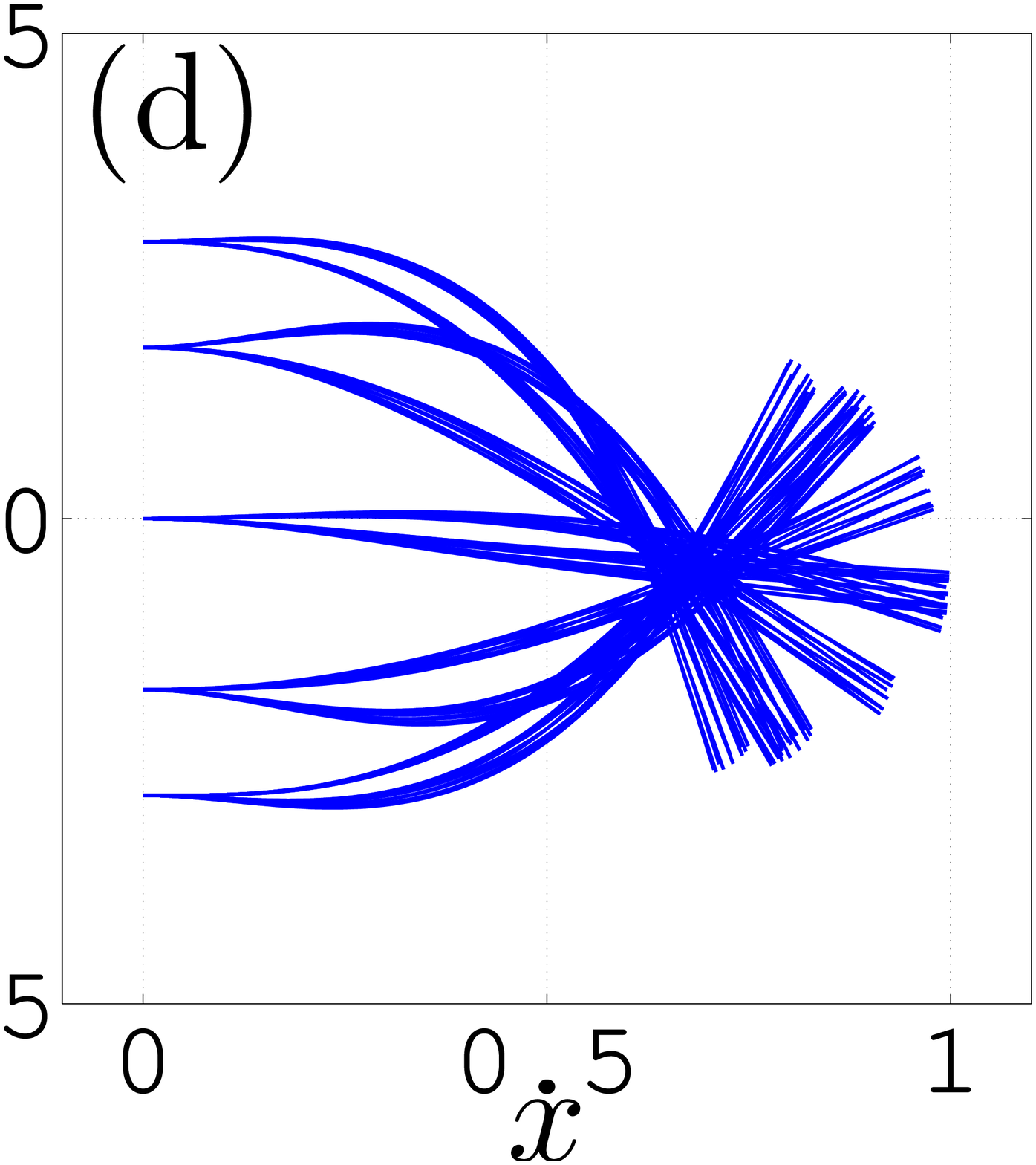}{1.2}  & \Dfig{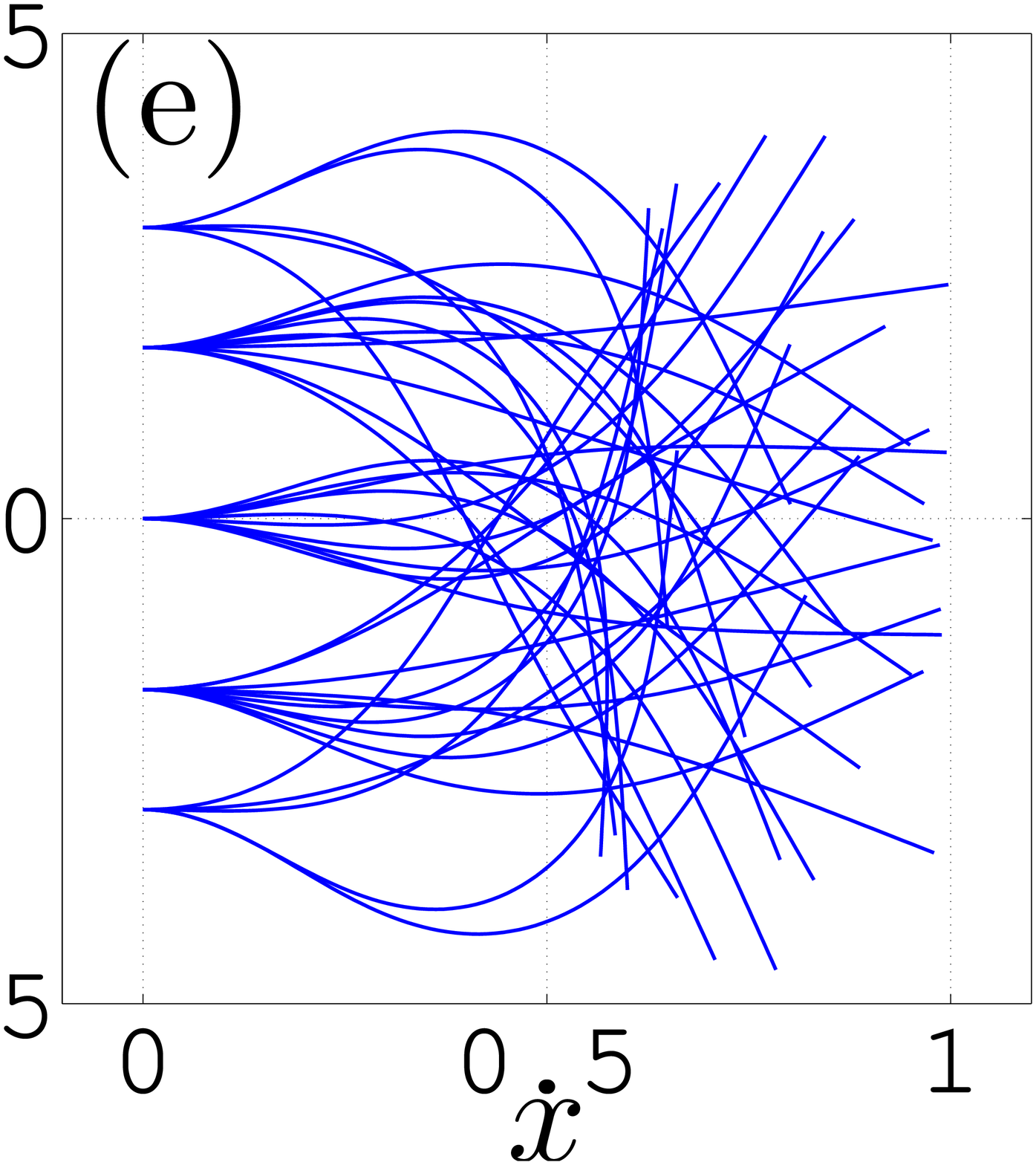}{1.2}\\
\Dfig{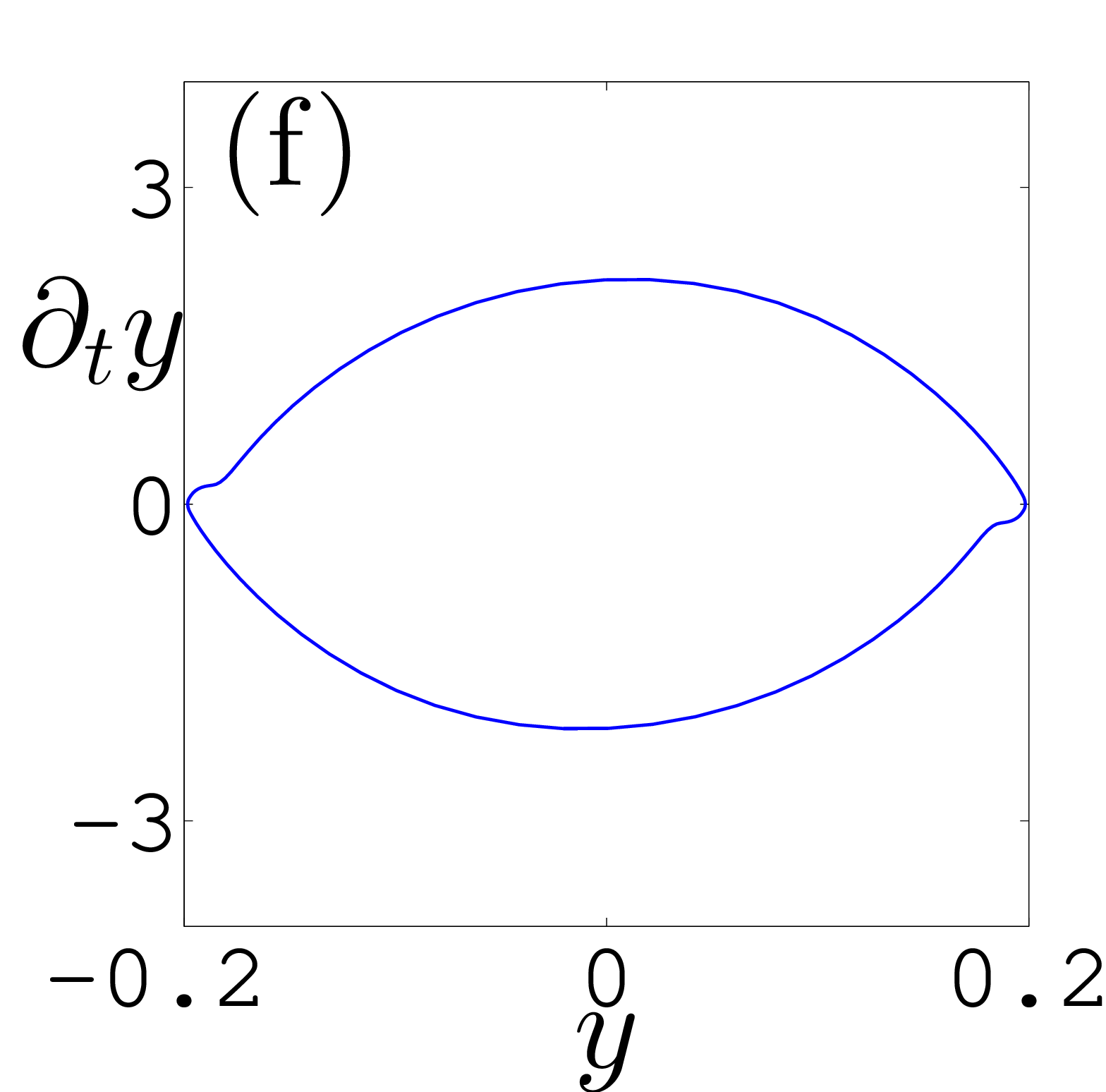}{1.2}& \Dfig{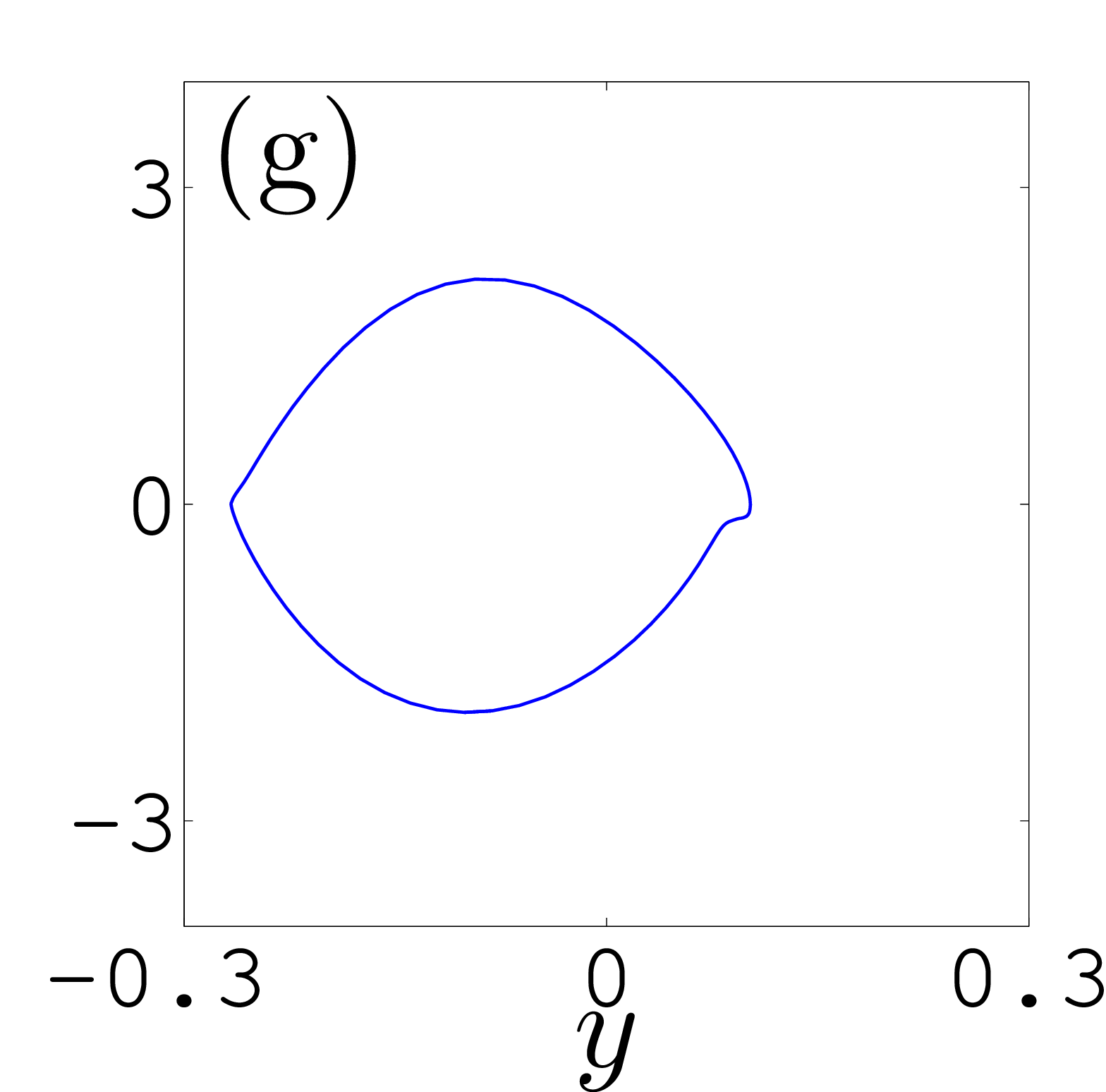}{1.2}& \Dfig{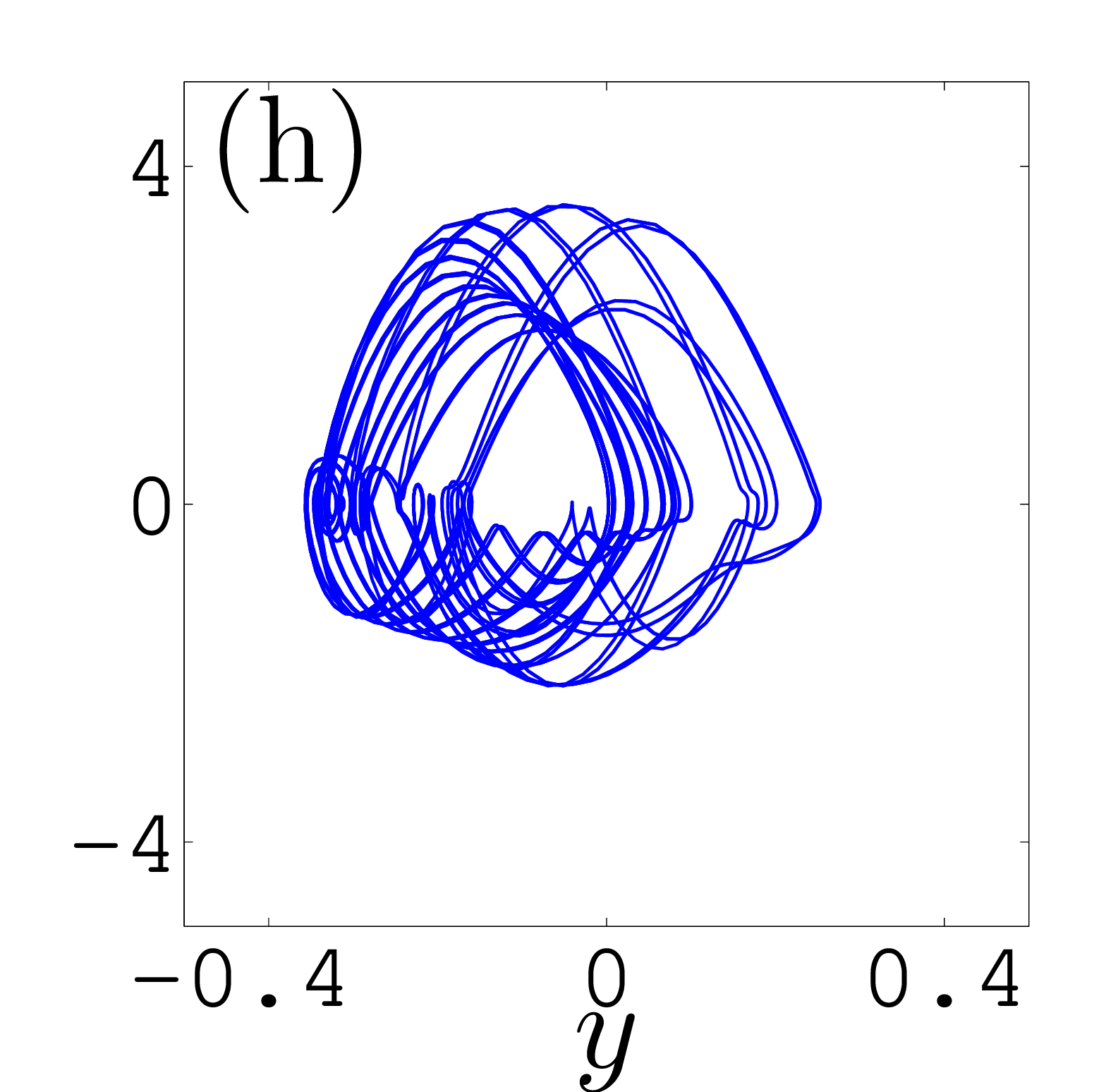}{1.2}& \Dfig{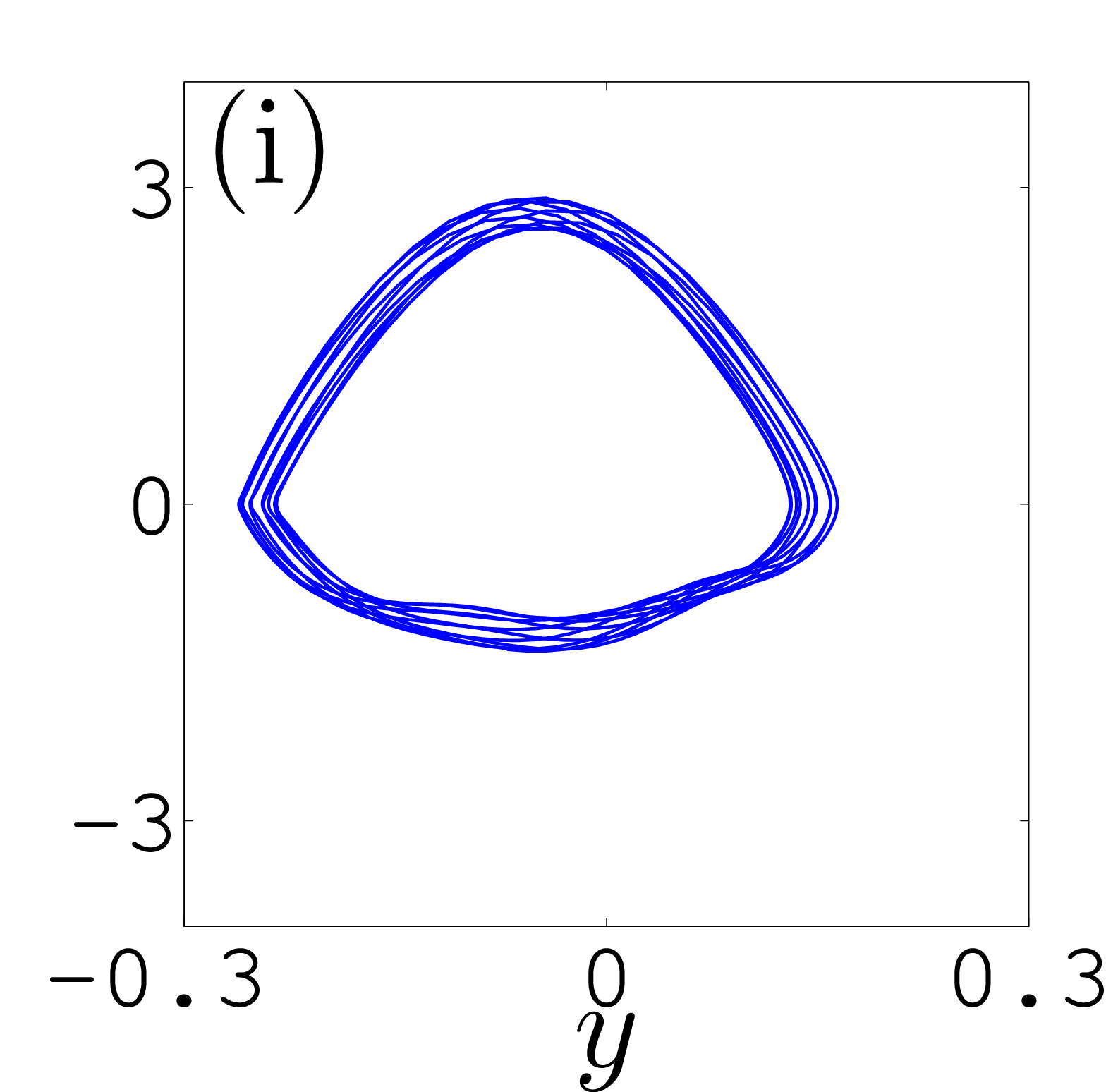}{1.2}& \Dfig{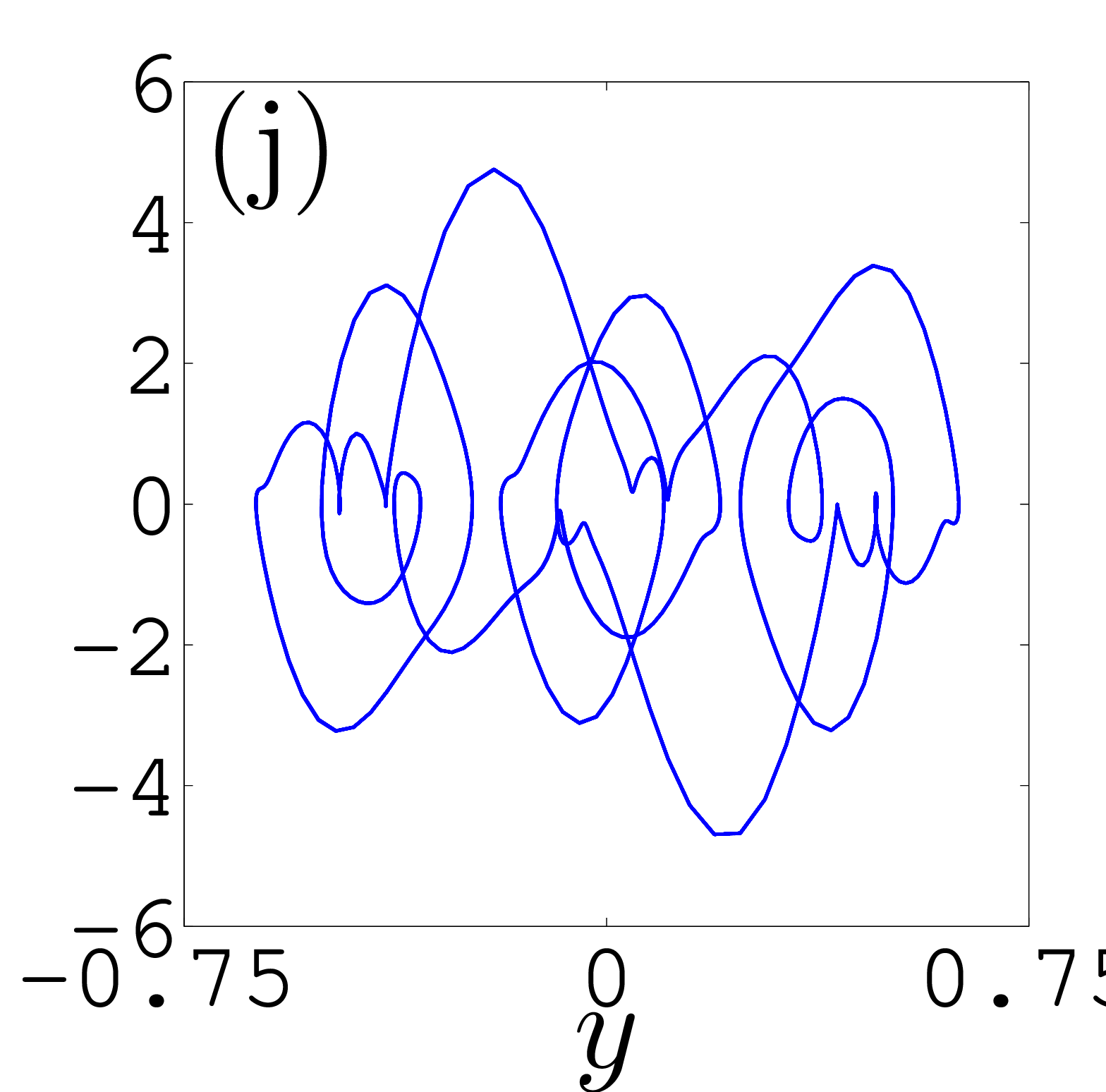}{1.2}\\
\rule{0pt}{1.5in}
\Dfig{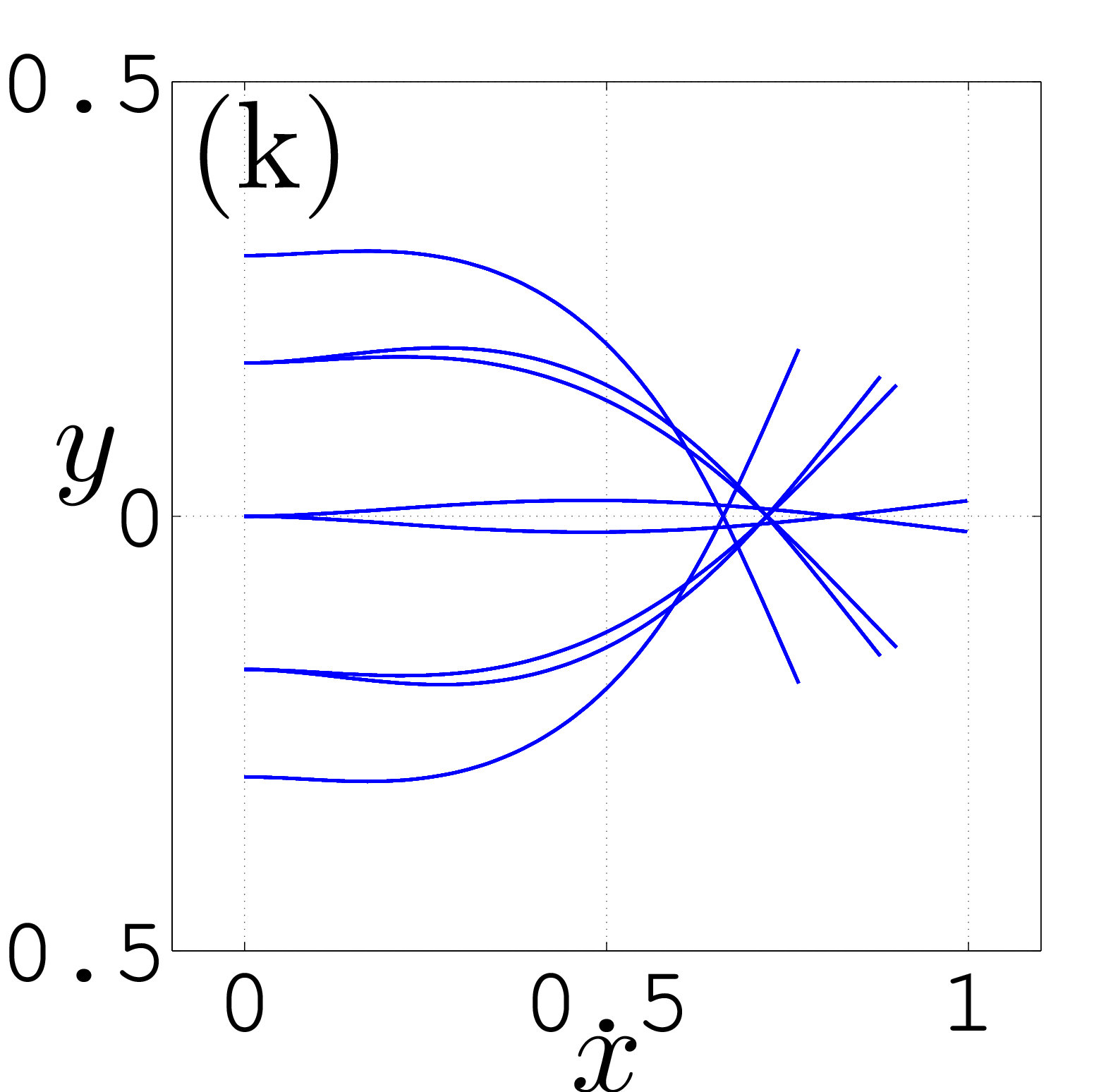}{1.2}& \Dfig{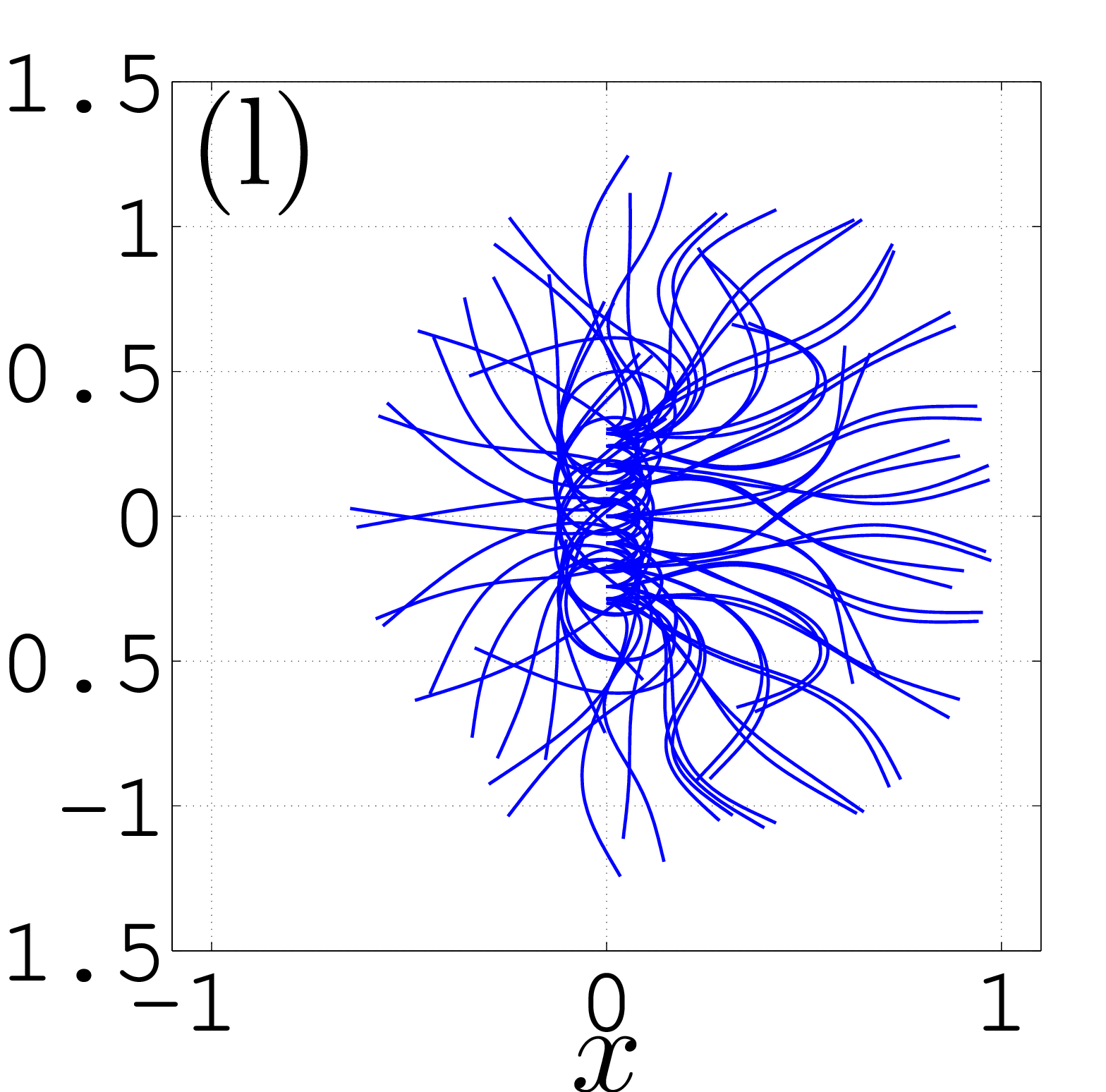}{1.2} & \Dfig{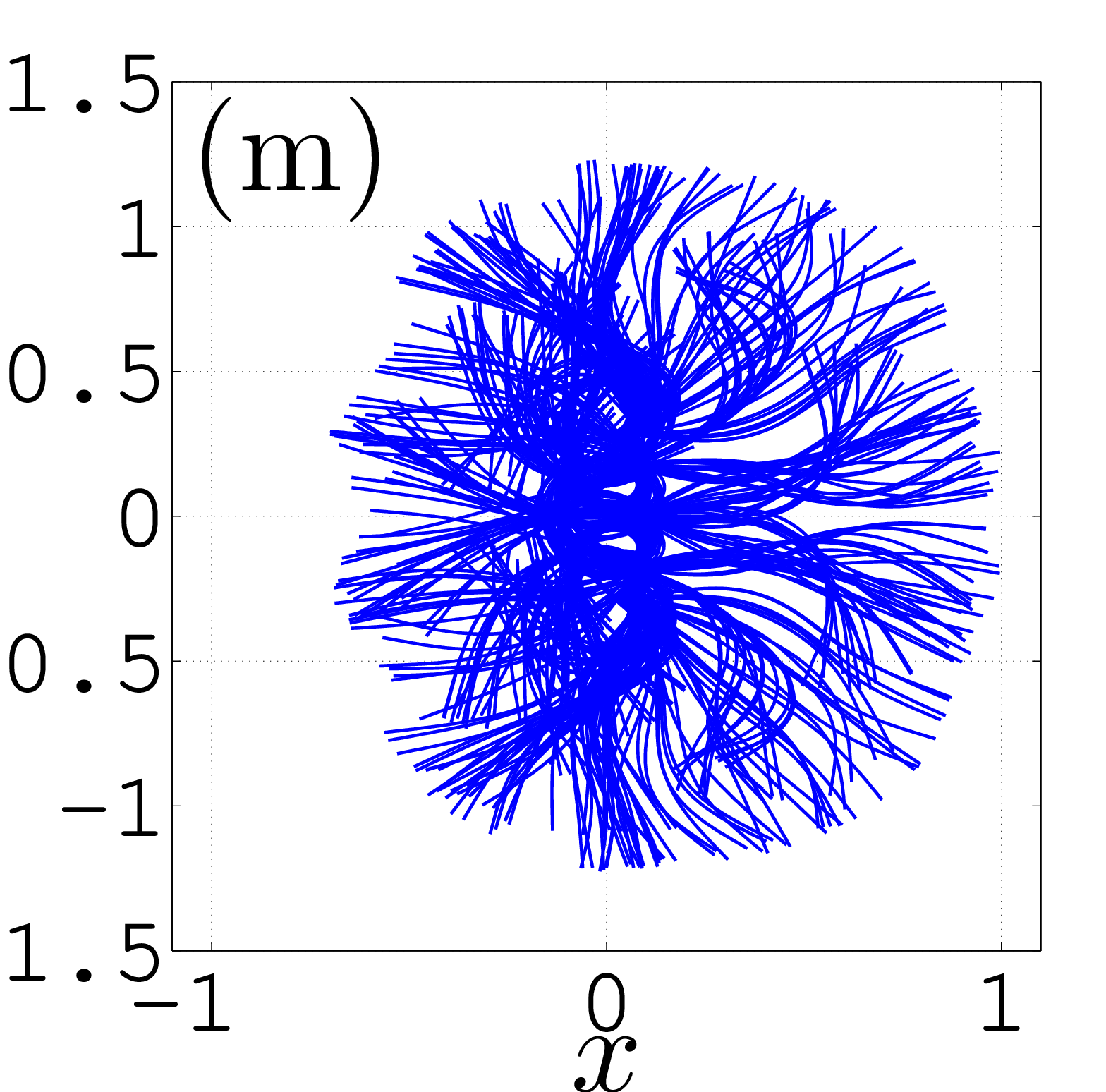}{1.2} & \Dfig{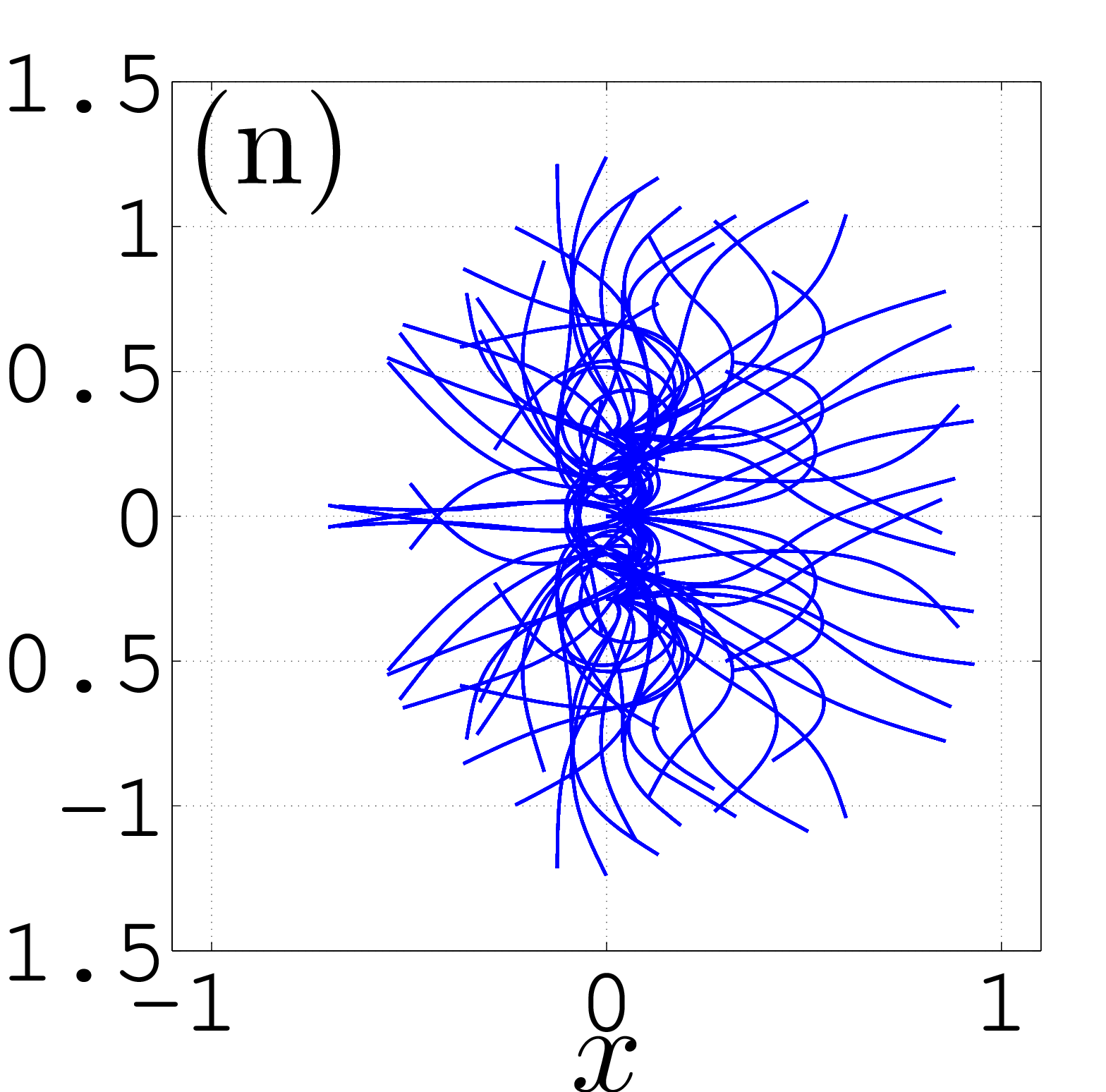}{1.2} & \Dfig{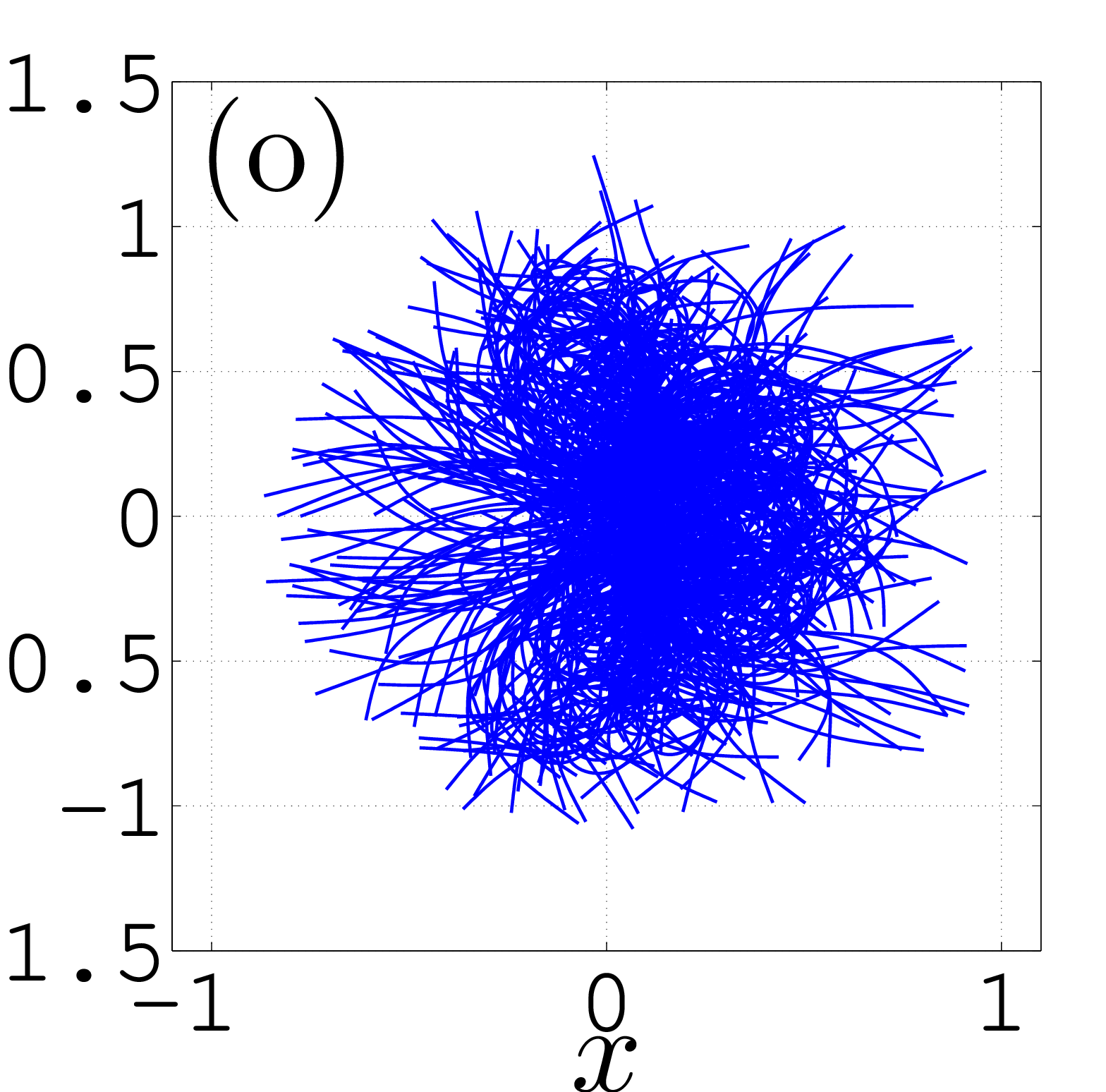}{1.2} \\
      \Dfig{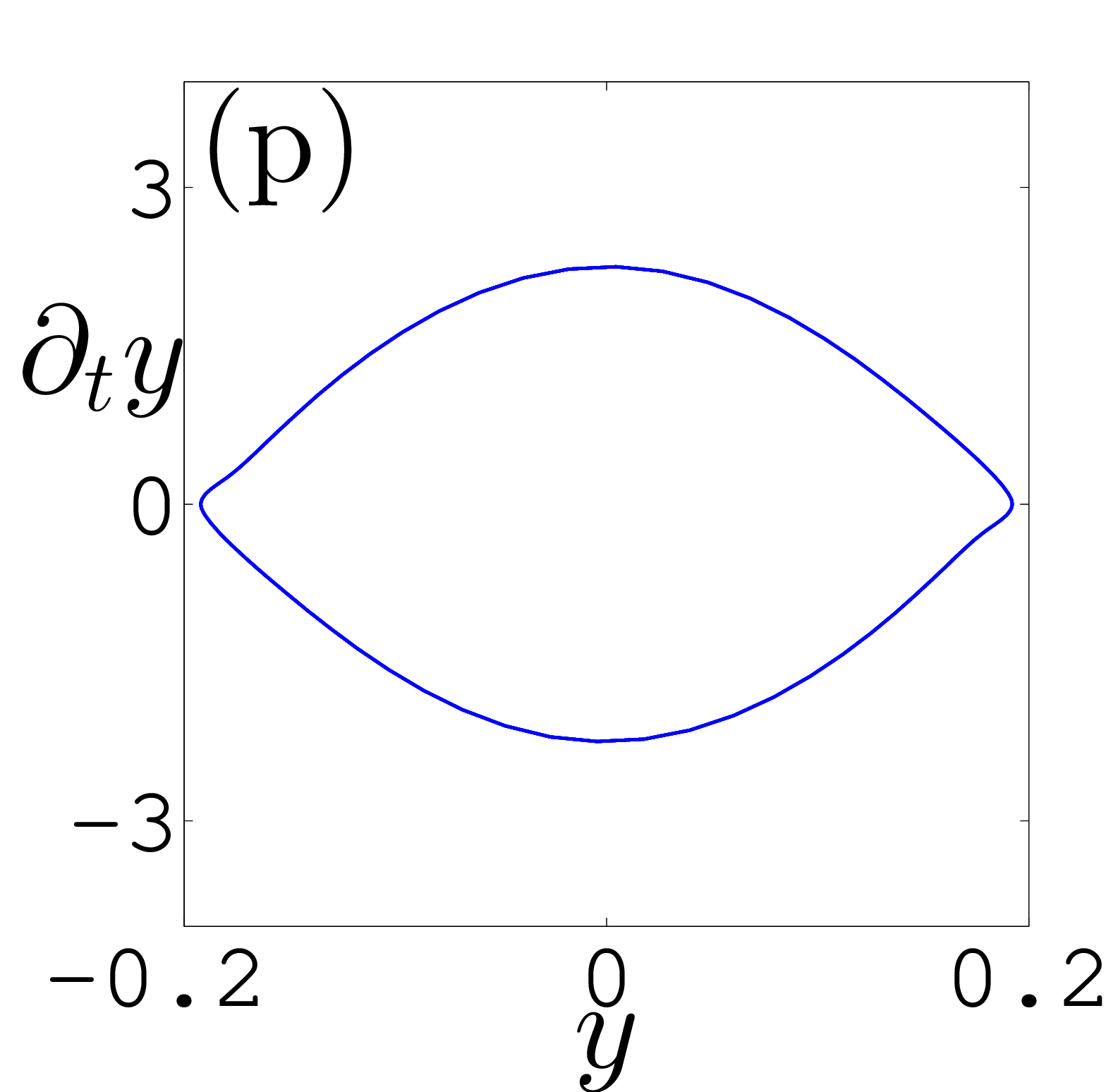}{1.2}&\Dfig{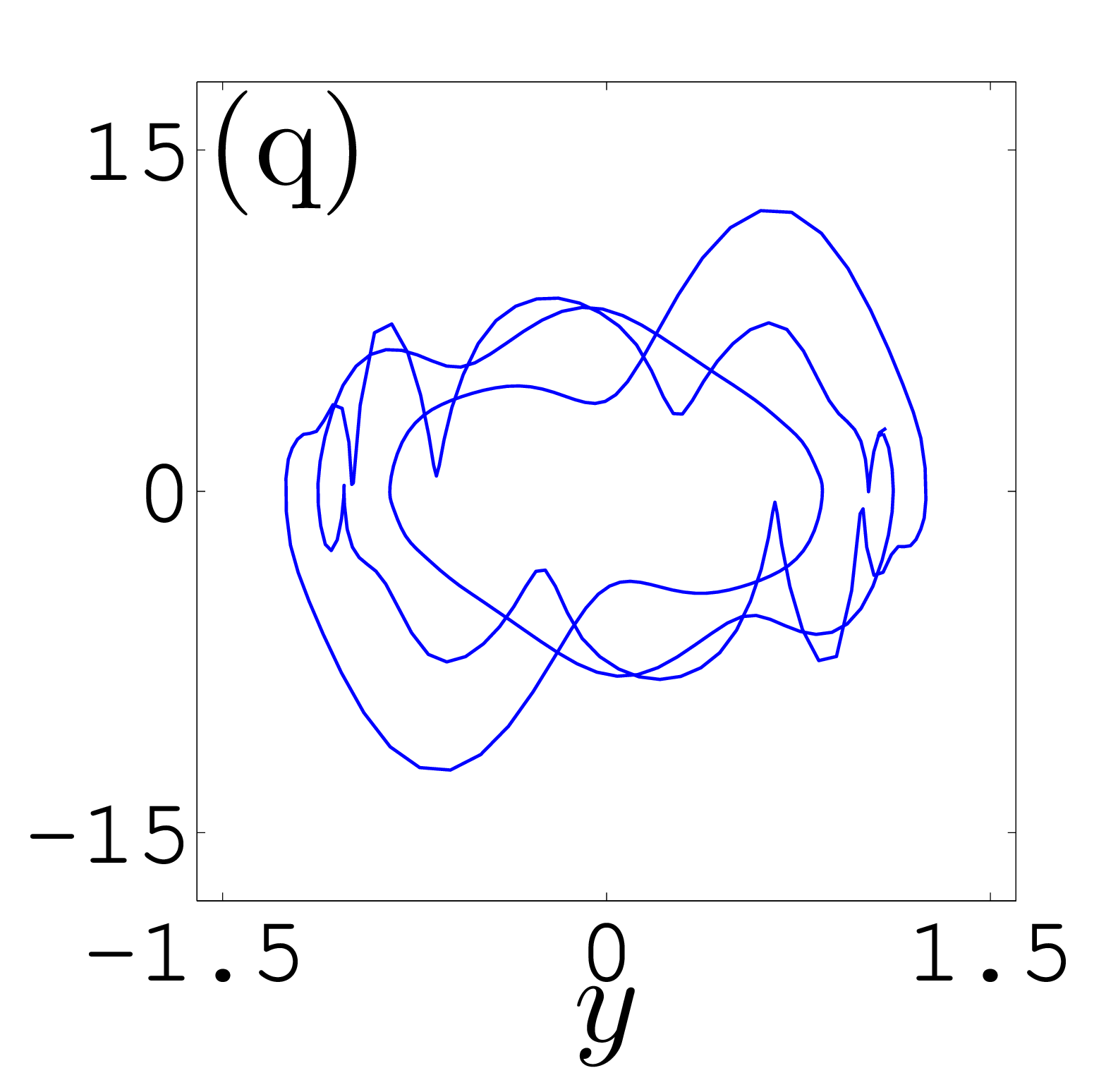}{1.2}& \Dfig{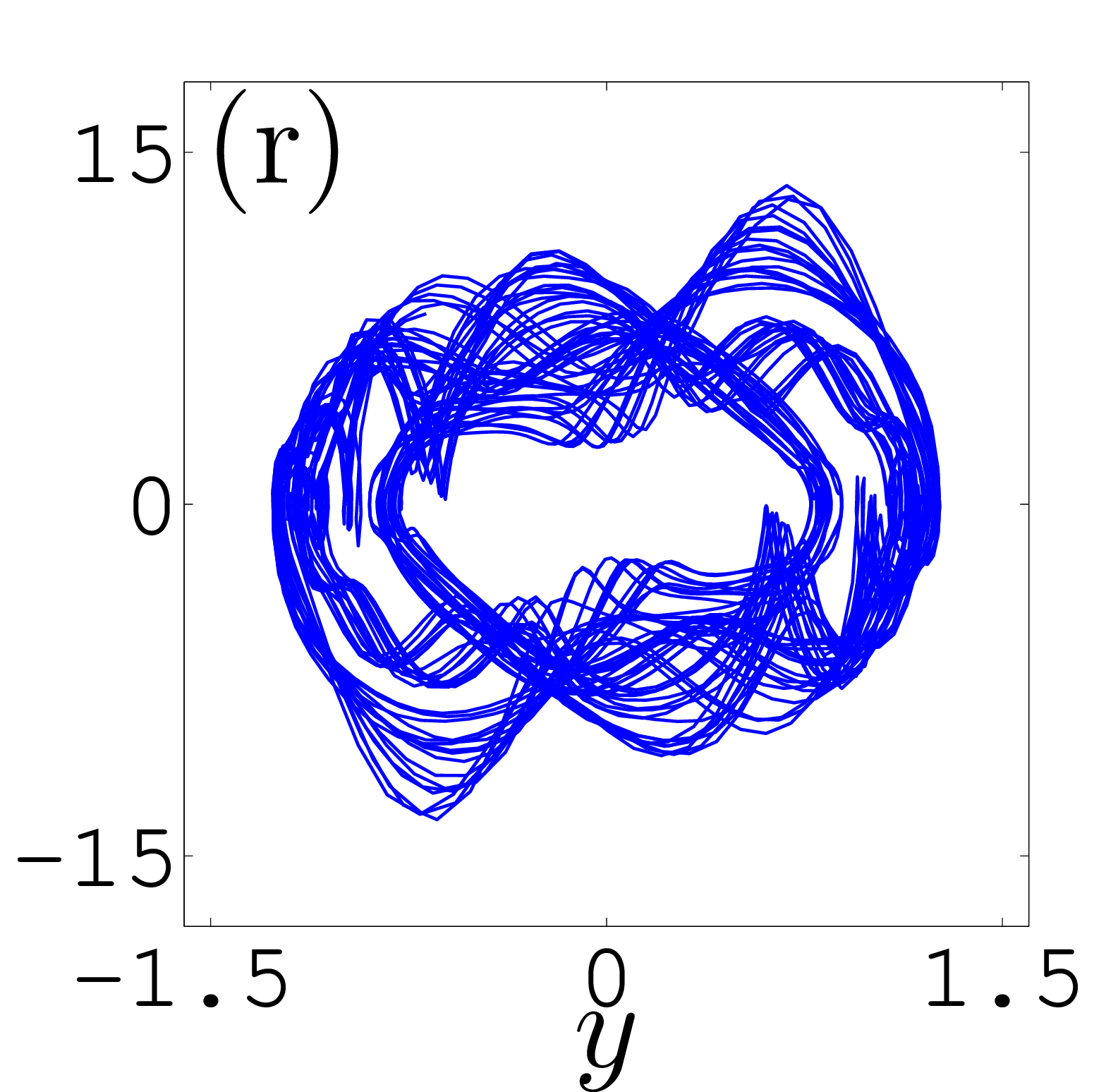}{1.2}& \Dfig{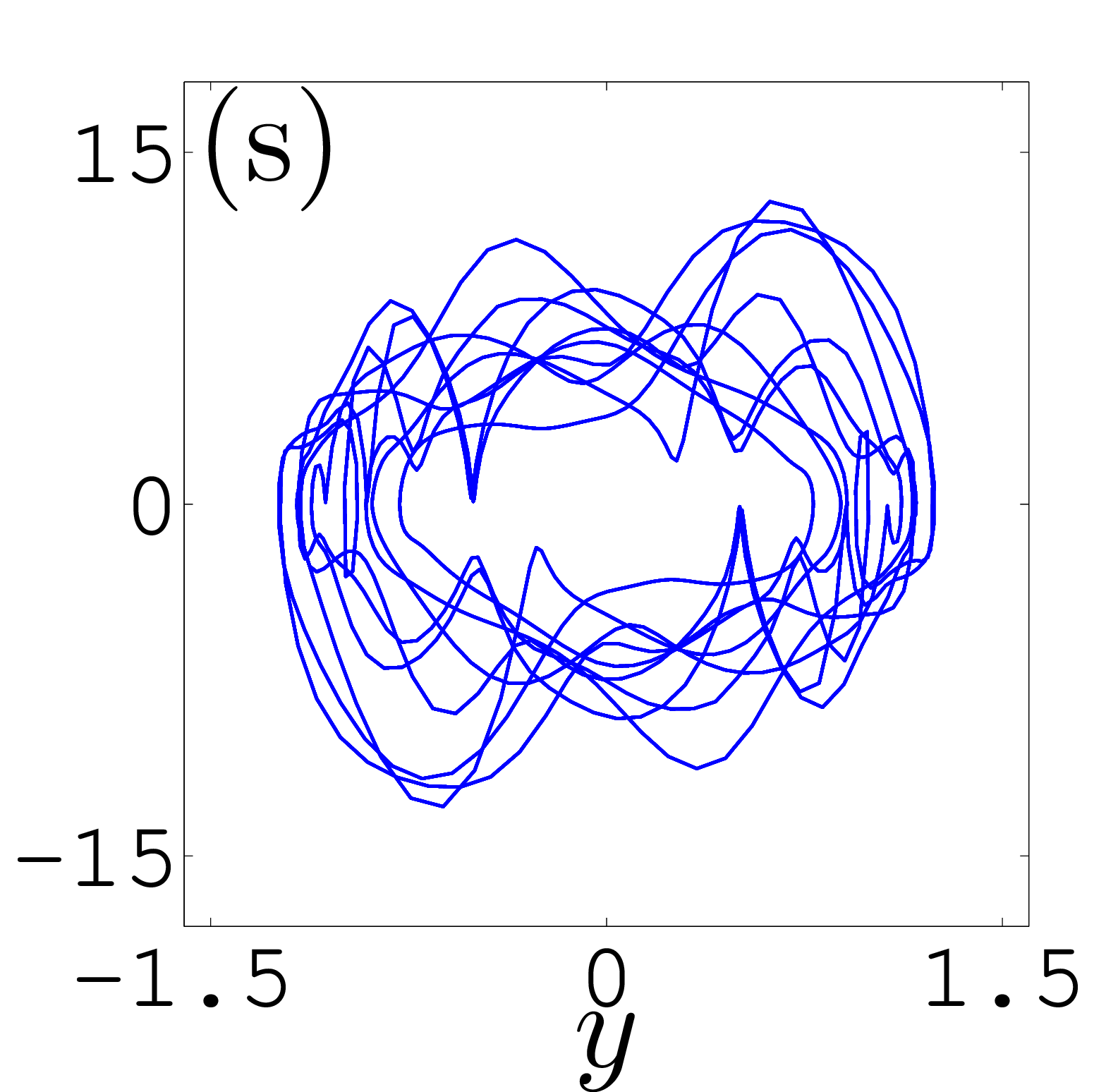}{1.2}& \Dfig{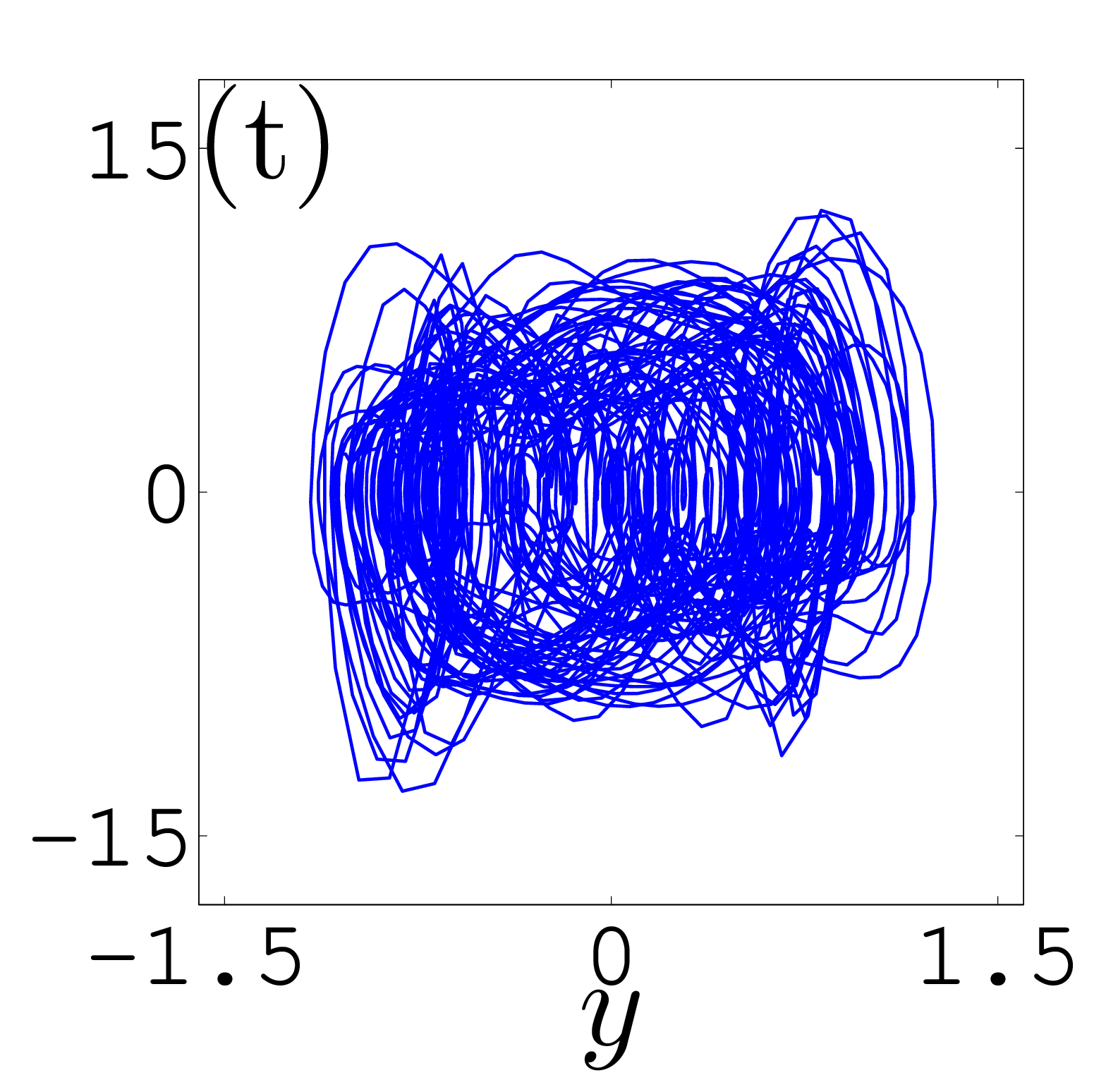}{1.2}
          \end{tabular}
          \caption{{\footnotesize\label{fig:periodorbit}}  Snapshots of the foil and phase plots of the free end vertical velocity versus
displacement in 50 periods, with $A=0.3$, $\mu=1$ and decreasing $B$. (a), (f). $B=0.6$, symmetric vibration of period 1;
(b), (g). $B=0.5$, asymmetric vibration of period 1; (c), (h). $B=0.48$, asymmetric vibration of period 13; (d), (i). $B=0.45$, asymmetric vibration of period 10;
(e), (j). $B=0.4$, asymmetric vibration of period 5; (k), (p). $B=0.35$, symmetric vibration of period 1; (l), (q). $B=0.33$, symmetric vibration of period 3;
(m), (r). $B=0.31$, non-periodic vibration; (n), (s). $B=0.3$, symmetric vibration of period 9;  (m), (t). $B=0.2$, non-periodic vibration.}
          \end{center}
  \end{figure}

We analyze the transition region for $A=0.3$ and $\mu=1$ as an example. The complex behavior of the system is shown in figure \ref{fig:FreqMu1A0.3}. In panel (a),
 we plot both the positive and negative vibration amplitude $A_v$, i.e., the positive and negative local maxima of the vertical displacement of the free end within one period,
for over 50 periods. The power spectrum density versus $B$
for the free end displacement is shown in figure \ref{fig:FreqMu1A0.3}(b), where the power spectrum for each case is normalized by the
corresponding maximum $A_v$. In figure \ref{fig:periodorbit}, we choose an example from each periodic and non-periodic state, and plot the phase plot of
the free end velocity in the vertical direction $\partial_ty$ against
the vertical displacement $y$ and the corresponding snapshots of the foil in 50 periods.

\Etwofigs{ampMu1A03}{Freqmu1A03}{\label{fig:FreqMu1A0.3}(a). Amplitude of the vibration $A_v$ in
50 periods with $A=0.3$, $\mu=1$ and various $B$. The different states include (1) symmetric vibration of period 1; (2) asymmetric vibration of period 1;
(3) asymmetric vibration of period 13; (4) asymmetric vibration of period 10; (5) asymmetric vibration of period 5; (6) symmetric vibration of period 1;
(7) symmetric vibration of period 3;
(8) non-periodic vibration; (9) symmetric vibration of period 9; (10) non-periodic vibration.
(b). The power spectrum density for the free end displacement. The results are normalized by the corresponding maximum $A_v$ for each $B$ value. }{3}{3}

As shown in figures \ref{fig:FreqMu1A0.3} and \ref{fig:periodorbit}, when the foil is rigid (large $B$), the vibration is
periodic and symmetric (region (1) in figure \ref{fig:FreqMu1A0.3}) with only a change in the amplitude. As $B$ decreases, the symmetry of the vibration about $x$-axis is broken. Depending on the initial condition,
the free end can move more to negative or positive $y$. The vibration goes through several different asymmetric periodic states as $B$ continues decreasing,
with
period 1 (region (2)),  period 13 (region (3)), period 10 (region (4)) and
period 5 (region (5)), and becomes symmetric again with period 1 in region (6). These six different periodic states can be viewed as a perturbation to
the symmetric period 1 vibration, as shown by the shapes of six foils in figure \ref{fig:periodorbit}, panels (a)-(e) and (k). When $B$ becomes even smaller, the foil is flexible enough to
allow more complex deformations and can flip over and vibrate in the left half plane ($x<0$). The vibration goes through several
periodic (region (7) with period 3 and region (9) with period 9) and non-periodic states (region (8)) until it stays non-periodic and chaotic in region (10) and beyond.
The dynamics of the foils become qualitatively different from those in regions (1) - (6).

When $\mu$ becomes even larger ($>7$ for $A=0.3$), the transition region disappears again. The vibration changes from period 1 for a more rigid foil to
non-periodic for a flexible foil directly, as shown in figure \ref{fig:periodiag}(b). However, at large $\mu$ and large $B$, the vibration stays asymmetric, which
is different from the symmetric motions at moderate $\mu$ and large $B$.

Next, we consider the effect of $A$ on the periodicity of the vibration. We fix the value of $B$ to be 0.5 and plot the diagram of different states of vibration
after 200 periods with various $A$ and $\mu$ in figure \ref{fig:periodiagA}. When $\mu$ is small, moderate, or large, the system exhibits different
dynamical behaviors as $A$ increases.
\Etwofigs{periodiagA}{periodiagAlargemu}{\label{fig:periodiagA}Diagram of periodicity of the vibration for $B=0.5$ with various $A$ and $\mu$.
 $\circ$: non-periodic vibration; $+$: symmetric periodic vibration of period 1;  $\triangleright$:
asymmetric vibration of period 1; $\times$: asymmetric periodic vibration of period 3.
 $\star$: symmetric vibration of period 7; $\bigtriangledown$: symmetric vibration of period 9; $\triangleleft$: symmetric vibration of
period 10; $\ast$: symmetric vibration of period 15; $\bullet$: symmetric vibration of period 17.}{3}{3}
When $\mu$ is small, the vibration transitions from periodic to non-periodic directly as the heaving amplitude increases. The critical $A$ is almost invariant
for different $\mu<1$. For moderate $\mu$, a transition region with different periods and symmetry is observed as $A$ increases. When $\mu$ is large enough, only
period 1 vibration is observed in the periodic regime. However, the foil becomes asymmetric first before it becomes non-periodic.

We note that similarly complicated dynamical behavior was also observed in flapping foil locomotion in a fluid medium. Chen
\etal identified five different periodic states and three chaotic states when a flag transitions from a periodic flapping to a non-periodic state under flow-induced vibrations \cite{chen2014bifurcation}. In \cite{spagnolie2010surprising}, Spagnolie \etal studied flapping locomotion with passive pitching in a viscous fluids, and observed a bistable regime where the wing can move either forward or
backward depending on its history. An asymmetric pitching motion was also observed in their work. In different systems, the dynamical behavior depends on different physical parameters including the driving frequency and amplitude, Reynolds number, etc. In our work, reduced heaving amplitude, frictional coefficients and bending rigidity are the most important parameters to consider.

\subsection{Resonance}
When the heaving frequency matches one of the natural frequencies of the vibration,
resonance occurs and the amplitude of the oscillation grows linearly with $t$.
In the nondimensionalized model, the heaving period is always 1. In the linearized model, the natural frequency depends on the rigidity of the beam $B$.
Therefore, a resonance occurs when
\bq
B=\displaystyle\frac{4\pi^2}{\omega_i^4},\quad i=1,2,\ldots \label{eq:resonanceB}
\eq
where $\omega_i$ are shown in Appendix B.
With the nonlinearities introduced by nonzero frictional forces and larger heaving amplitude $A$, the resonant $B$ values also vary accordingly.

We first consider the effect of the frictional force on the resonance. In figure \ref{fig:resonancemu}(a), we plot the free end amplitude $A_v$ versus the foil rigidity $B$ for a fixed heaving amplitude $A=0.05$ and various $\mu$. We plot the free end amplitude in the linearized model with
no friction with a dashed line.
\Efivefigsv{resonancemu}{B6_mut01_A005mode}{B1_mut01_A005mode}{B009_mut01_A005mode}{B002_mut01_A005mode}{\label{fig:resonancemu}
(a). Vibration amplitude $A_v$ vs. foil rigidity $B$ for fixed $A=0.05$ and various $\mu=0.1,0.5$ and 1. The linearized analytical solution with $\mu=0$ is denoted
by the dashed line; (b), first vibration mode with $A=0.05$, $\mu=0.1$ and $B=6$;
(c), second vibration mode with $A=0.05$, $\mu=0.1$ and $B=1$; (d), third vibration mode with $A=0.05$, $\mu=0.1$ and $B=0.09$.
(e), fourth vibration mode with $A=0.05$, $\mu=0.1$ and $B=0.02$.}{3}{1.2}
As $B$ decreases, multiple resonances are observed with $\mu=0$ according to equation (\ref{eq:resonanceB}). We only plot the first three of the infinite sequence of resonances in the panel. As $\mu$ increases, the amplitude $A_v$ decreases correspondingly. The resonant $B$ values shift to the right as $\mu$ increases except for the first resonance. As $B$ decreases, the shape of the foil changes from the first mode to higher bending modes. In panels (b)-(e), we plot snapshots of the foil in one period for $A=0.05$, $\mu=0.1$ and $B=6$, 1, $0.09$ and $0.02$
respectively, showing the different modes.

\Efourfigs{resonanceA}{resonanceA_Mu1}{modeB30mu01A03}{modebisB30mu01A03}{\label{fig:resonanceA}
(a). Vibration amplitude $A_v$ vs. foil rigidity $B$ with $\mu=0.1$ and various $A=0.03, 0.05, 0.07,0.1,0.3$ and 0.5;
(b). $A_v$ vs. $B$ zoomed in near the bistability with the same parameter.
(c). snapshots of foil in one period with $A=0.3$, $\mu=0.1$ and $B=3.05$ in lower branch; (d). snapshots of foil in one period with $A=0.3$, $\mu=0.1$ and $B=3.05$ in upper branch.  }{2.7}{2.7}

Next, we consider the effect of $A$ on the resonance. In figure \ref{fig:resonanceA}(a), we plot the free end amplitude $A_v$ versus
the bending rigidity $B$ with fixed $\mu=0.1$ and various $A$. We only consider symmetric vibrations with period 1 in this figure; the curves
in panel (a) end when the vibration becomes either non-periodic or asymmetric for the particular parameter values.

When $A$ is small ($A<0.1$ in this case), the free end amplitude $A_v$ increases with increasing $A$. Moreover, the resonant $B$ value shifts to the right as $A$
becomes larger at the second ($B\approx 0.1$) and third peaks ($B\approx 0.01$). When $A$ becomes larger, we observe bistability
near the first resonant value ($B\approx 3$). In figure \ref{fig:resonanceA}(b), we enlarge the region near the first resonant value to show the bistability.

Typical motions in the bistable regime are shown in figures
 \ref{fig:resonanceA}(c) and (d), which
show the snapshots of the foil in one period for $\mu=0.1$, $A=0.3$ and $B=3.05$. The snapshots corresponding to the lower branch are shown in panel (c) and those
corresponding to the upper branch are shown in panel (d). For the lower branch, as the leading edge of the foil moves upward by the heaving motion, the trailing edge moves downwards. For the upper branch, the trailing edge moves in the same direction as the leading edge. This phenomenon is observed for both $A=0.3$ and $A=0.5$,
and as $A$ increases, the bistability region becomes larger.

The bistability is observed for different values of $\mu$ as long as $A$ is large enough. It is a result of the nonlinearity of the system. Such a double fold bifurcation near a
resonant peak has been observed in other nonlinear oscillators such as the Duffing oscillator \cite{wagg2016nonlinear}.
\section{Free Locomotion}
We now consider free locomotion of the foil: the leading edge can move freely along the $x$-direction with position $X_0(t)$. A periodic vertical heaving motion $ y(0,t)=A\sin(2\pi t)$
 is still applied at the
leading edge, and to solve for $X_0(t)$, we assume no tangential force (tension) is applied at the leading edge:
\bq
T(0,t)=0
\eq
The rest of the boundary conditions corresponding to the free motion at the leading and trailing edges are the same as in the fixed base case.

As the foil bends, a horizontal force is obtained from transverse friction, and we expect the foil to move horizontally in general.
We define the space and time averaged horizontal velocity as $\bar{u}=\displaystyle\int_0^1\int_0^1\partial_tx(s,t)dsdt$. We define the direction
to the right as positive, and observe that in general $\bar{u}$ takes negative values. The
only input to the system is the leading edge heaving, and thus the input power can be evaluated by the power applied at the leading edge
 $\bar{P}=\displaystyle\int_0^1\partial_ty(0,t)B\partial_s\kappa(0,t)dt$, where $\partial_ty$ is the vertical velocity component and $B\partial_s\kappa$ is the
shearing force in the vertical direction at the leading edge. In figure \ref{fig:SwimBvsMut}, we plot $-\bar{u}$ and $\bar{P}$ versus $B$ for fixed $\mu_f=0.01$, $A$=0.1,
and various $\mu_t$ such that $\mu_f \ll \mu_t$. When $\mu_t$ is small (less than 5), a resonant peak is obtained in $\bar{P}$ near $B\approx2$ and corresponds to a decrease in the
velocity. At larger $\mu_f$ (not shown), the foil has a smaller horizontal speed (unsurprisingly), and stronger resonant-like behaviors. Both features are present in the lower curves in Figure \ref{fig:SwimBvsMut}(a), and these features are strengthened as $\mu_f$ increases. For a real snake, $\mu_f<\mu_t$ but both are in the range 1-2 \cite{Hu09,MaHu2012a,HuSh2012a}. In this regime, our passive elastic foil translates slowly ($\approx0.1$ body lengths per period), and for certain values of $B$ moves rightward (toward the free end) at large amplitudes ($A > 0.1$). Due to the slow speed of locomotion, the foil behavior has strong similarities to the fixed base case.

The upper curves in figure \ref{fig:SwimBvsMut}(a) ($\mu_t\geq 5$) tend towards the case of wheeled robots with large transverse friction and  small tangential friction \cite{transeth09}, where the foil has a higher speed and efficiency.
\Etwofigs{SwimSpeedBvsMutMuf001}{SwimPowerBvsMutMuf001}{\label{fig:SwimBvsMut} (a). Negative horizontal velocity $-\bar{u}$ vs. $B$ for fixed $\mu_f=0.01$, $A$=0.1 and various $\mu_t$
= 0.3, 1, 3, 5, 10, 30, 100. (b). input power $\bar{P}$ vs. $B$ for the same $\mu_t$.}{2.7}{2.7}
We show how the foil motion changes from small to large $\mu_t$ in figures \ref{fig:SwimMode}(a), (b) and (c). We plot snapshots of the foil as well as the trajectory
of the leading edge in one period with
fixed $A=0.1$ and $\mu_f=0.01$, and various $\mu_t$=1, 10, 100 and $B$=0.5, 2.5, and 10.
\Efourfigs{SwimModeMut1A01}{SwimModeMut10A01}{SwimModeMut100A01}{slipcontour}{\label{fig:SwimMode}
Snapshots of foil and the trajectory of the leading edge in one period with fixed $A=0.1$, $\mu_f=0.01$, and various $B=0.5,2.5,10$, and (a). $\mu_t=1$;
(b). $\mu_t=10$; (c). $\mu_t=100$. Another snapshots are plotted with all the leading $x$ position relocated at the same value.
(d). Contour plots of $y(x,t)$ over $x$ and $t$ within one period for $A=0.1$, $B=10$, $\mu_f=0.01$ and $\mu_t=100$. }{2.7}{2.7}
At the right of each panel, we relocate the foil so that the snapshots at different time instants share the same
leading $x$ location, to better illustrate the mode shapes. In panel (a), $\mu_t=1$, and the foil shows two different mode shapes for $B=0.5$ and $10$. Near the resonant peak at $B\approx 2.5$, the foil vibrates
in a large-amplitude motion mainly along the transverse direction. There, the $x$ velocity decreases significantly while the input power increases greatly.
As $\mu_t$ increases to 10, the foil transitions to a different dynamical regime, where the velocity and the input power vary more smoothly and the resonant peaks are reduced as
shown in figure \ref{fig:SwimBvsMut}. In figure \ref{fig:SwimMode}(b), the differences in the mode shapes are reduced compared to panel (a).
 When $\mu_t = 100$, the foil deflection is much smaller as shown in figure
\ref{fig:SwimMode}(c). When $\mu_t$ is large enough, transverse friction
dominates the system, while the foil is close to a flat plate as $B\rightarrow\infty$. A fully rigid plate is horizontal, so transverse friction provides no thrust force. Therefore, the horizontal velocity $\bar{u}$ is zero in this limit. In
figure \ref{fig:SwimBvsMut}(a), we see that $\bar{u}\rightarrow 0$ as $B$ becomes larger, and a maximum speed is obtained at a moderate $B$ value.
In
figure \ref{fig:SwimBvsMut}(b), we observe that the input power $\bar{P}$ approaches a certain limit as $B$ increases for a fixed value of $\mu_t$, since the foil converges to a purely vertical oscillation and the work done against transverse friction is independent of $B$ for a rigid plate.

Figure \ref{fig:SwimMode}(c) indicates that in the large $\mu_t$ regime, the motion of the foil can be approximated by a travelling wave solution $g(x-U_wt)$ where
$U_w$ is a wave speed, as the snapshots of the foil seem to follow a certain wave track. In previous work \cite{alben2013optimizing,wang2014optimizing} we found that
a travelling wave motion was optimal for efficiency at large $\mu_t$. In the current model, we do not prescribe the shape of the foil, so it is interesting that at large $\mu_t$, the flexible foil spontaneously adopts a travelling wave motion.
To clearly illustrate
the travelling wave motion, we show a contour plot of $y(x,t)$ versus $x$ and $t$ within one period in figure \ref{fig:SwimMode}(d), for $A=0.1$, $B=10$, $\mu_f=0.01$ and $\mu_t=100$.
For $x$ away from the leading edge, we find that the contour curves are close to straight lines, so $y$ is of travelling wave form. Deviations are
observed for $t$ near 0.25 and 0.75, when $y$ reaches is extrema. This is reasonable because the velocity of $y$ changes sign at its extrema and the deflection of the foil cannot be characterized as a traveling wave there. We also note that at the leading edge, the foil is held flat ($\partial_x y=0$) at all times, while the travelling wave has nonzero slope $\partial_x y\neq 0$. This can be seen in how the contours in Figure \ref{fig:SwimMode}(d) change from zero slope at $x = 0$ to nonzero slope $(= U_w)$ for $0.5 \lesssim x \leq 1$.
To satisfy the clamped boundary condition, we expect (and find) a boundary layer at the leading edge, as we now describe.

Along with a boundary layer form, the approximate traveling wave solutions at large $\mu_t$ also obey certain scaling laws. Two of the most important quantities are the horizontal speed $-\bar{u}$ and the input power $\bar{P}$. In figure \ref{fig:mutscaling}(a), we show that $-\bar{u}\sim \mu_t^{1/4}$ and in figure
\ref{fig:mutscaling}(b), $\bar{P}\sim\mu_t^{5/12}$. By assuming small slopes ($|\partial_x y|\ll1$) and approximate traveling wave solutions outside of a boundary layer, we now explain these scaling laws.

\Etwofigs{uvsmutscaling}{pvsmutscaling}{\label{fig:mutscaling} For $A=0.1$, $\mu_f=0.01$ and $B= 1,5,10,100$, (a). $\log_{10}\mu_t$ vs. $\log_{10}(-\bar{u})$. A
$\mu_t^{1/4}$ is scaling law is shown with a dashed line. (b). $\log_{10}\mu_t$ vs. $\log_{10}(\bar{P})$. A $\mu_t^{5/12}$ scaling law is shown with a dashed line.}{3}{3}

When the amplitude $A$ is small, we can simplify the foil model in the large limit of $\mu_t$ by approximating the
arclength $s$ by $x$, and $\zeta(s,t)\approx x+iy(x,t)$. At leading order, the tangent and normal vectors are:
\begin{equation}
 \hs\approx(1,\partial_xy),\quad \hn\approx(-\partial_xy,1).
\end{equation}
The horizontal velocity $\partial_t x$ has small variations over arclength and time, and therefore $\partial_tx(s,t)\approx \bar{u}\equiv U$,
and $\partial_t\zeta(x,t)\approx U+i\partial_t y$.
In the limit of large $\mu_t$, $U\gg \partial_t y$, as we expect the vertical velocity $\partial_ty$ is proportional to the heaving amplitude $A$ while $U$ grows with $\mu_t$.
Therefore, the normalized velocity at leading order is $\hptZ\approx (-1,-\displaystyle\frac{\partial_ty}{U})$. The negative sign is obtained as we only consider the case
where the body moves to the left ($U<0$). The normalized tangential and normal velocity components are, to leading order,
\begin{equation}
 \hptZ\cdot\hs\approx -1,\quad \hptZ\cdot\hn\approx \partial_xy-\displaystyle\frac{\partial_ty}{U}.
\end{equation}
Now, we take the $y$-component of equation (\ref{eq:nonbeam}), neglect higher order terms,
and obtain a force balance for the vertical motion:
\bq
\partial_{tt}y=-B\partial_x^4y-\mu_t\left(\partial_xy-\displaystyle\frac{\partial_ty}{U}\right)\label{eq:simbeam}
\eq
where $\kappa\approx\partial_x^2y,\partial_s\kappa\approx\partial_x^3 y$ and $\partial_{ss}\kappa\approx\partial_x^4y$.

As mentioned already, the leading edge clamped boundary condition is not compatible with a travelling wave solution, so we look for a boundary layer near the leading edge.
\Ethreefigs{dxy1}{dxybl}{dx4ybl}{\label{fig:boundarylayer}For a fixed $B=10$, $A=0.1$ and $\mu_f=0.01$ and various $\mu_t$
(a). $x$ vs. $\partial_xy$; (b). $x\mu_t^{1/3}$ vs. $\partial_xy\sim\mu_t^{1/4}$; (c). $x\mu_t^{1/3}$ vs. $\partial_x^4y\sim\mu_t^{-3/4}$. Good collapse of curves is
obtained within the boundary layer, which indicates the length of the boundary layer scales as $\mu_t^{-1/3}$, $\partial_xy\sim O(\mu_t^{-1/4})$,
and $\partial_x^4y\sim O(\mu_t^{3/4})$. }{1.9}{1.9}{1.9}
In figure \ref{fig:boundarylayer}(a), we plot $\partial_xy$ versus $x$ for $B=10$, $A=0.1$ and $\mu_f=0.01$ and various $\mu_t$. We find
that $\partial_xy$ scales as $\mu_t^{-1/4}$ from the numerical simulations.
Since $\partial_xy=0$ when $x=0$, $\partial_xy$ will increase from 0 to $O(\mu_t^{-1/4})$ within the
boundary layer. The vertical velocity $\partial_t y$ and acceleration $\partial_{tt}y$ are $O(1)$ $(\sim A)$ near the leading edge. Thus, according to equation (\ref{eq:simbeam}), we have the following scalings within the boundary layer:
\bq
\mu_t\left(\partial_xy-\displaystyle\frac{\partial_ty}{U}\right)\sim\mu_t(-1/U)\sim\mu_t^{3/4}\Rightarrow \partial_x^4y\sim\mu_t^{3/4}
\eq
We pose the boundary layer width as $\mu_t^\alpha$. Using $\partial_xy\sim\mu_t^{-1/4}$ and $\partial_x^4y\sim\mu_t^{3/4}$,
and assuming that each differentiation divides by a factor proportional to the boundary layer width, we have that $\alpha=-1/3$, and the length
of the boundary layer scales as $\mu_t^{-1/3}$. In figures \ref{fig:boundarylayer}(b) and (c), we plot $\partial_xy$ scaled by $\mu_t^{1/4}$, and $\partial_x^4y$ scaled by $\mu_t^{-3/4}$ versus $x\mu_t^{1/3}$ and find a good collapse for both quantities within the boundary layer, particularly at larger $\mu_t$.

The time-averaged input power $\bar{P}=\displaystyle\int_0^1B\partial_x^3y(0,t)\partial_ty(0,t) dt$. The scaling of $\partial_x^3y$ is obtained by differentiating $\partial_xy \sim \mu_t^{-1/4}$ twice with respect to x. Each differentiation divides by a factor of $\mu_t^{-1/3}$, the boundary layer width. Consequently, $\partial_x^3 y \sim \mu_t^{-1/4 + 2/3} = \mu_t^{5/12}$. Since $\partial_t y \sim A \sim 1$, $\bar{P}$ also scales as $\mu_t^{5/12}$.  This scaling is confirmed by the numerical results
in figure \ref{fig:mutscaling}(b).

We briefly mention the dependence of the locomotion on the heaving amplitude $A$. For $\mu_t\lesssim 1$, we recall there are  resonant peaks
where the velocity significantly decreases and the input power increases.
\Efourfigs{SwimSpeedBvsAsmallmut}{SwimPowerBvsAsmallmut}{SwimSpeedBvsAlargemut}{SwimPowerBvsAlargemut}{\label{fig:SwimBvsA}
(a). Negative horizontal velocity $-\bar{u}$ vs. $B$ for fixed $\mu_f=0.01$, $\mu_t$=1 and various $A$
= 0.03,0.05,0.07,0.1,0.3,0.5. (b). clamped power $\bar{P}$ vs. $B$ for the corresponding parameters.
(c). $-\bar{u}$ vs. $B$ for fixed $\mu_f=0.01$, $\mu_t$=100 and various $A$. (d). $\bar{P}$ vs. $B$ for the corresponding parameters. }{2.7}{2.7}
In figure \ref{fig:SwimBvsA}(a), we plot the horizontal speed for $\mu_t = 1$. Near the first resonance ($B\approx3$), the speed has a local minimum, and as we increase $A$ from 0.03 to 0.1, the minimum does not move much, but the trough around the minimum broadens. Increasing $A$ further to 0.3 and 0.5, strong nonlinear effects come into play, and the foil reverses its horizontal direction slightly below the resonance. This rightward motion is shown in figure \ref{fig:SwimModeA}(a), at $B = 1.8$. For comparison, the leftward body motion at resonance is shown in panel (b) at $B = 2.5$. The two flapping modes are clearly quite different. For the reverse motion, the leading and trailing edges oscillate in opposite direction vertically, while for the other case,
they move in the same direction. This is consistent with the results we obtained in the fixed base cases (figure \ref{fig:resonanceA}(c) and (d)).
In figure \ref{fig:SwimBvsA}(b), we plot the input power for $\mu_t = 1$ near the first resonance. As $A$ increases, the resonant peak broadens and becomes less symmetrical, similarly to the fixed-base case.
\Etwofigs{SwimModeB18Mut1A03}{SwimModeB25Mut1A03}{\label{fig:SwimModeA}
Snapshots of the foil and the trajectory of the leading edge in one period with $A=0.3$, $\mu_f=0.01$, $\mu_t=1$,
 and (a). $B=1.8$, reverse motion;
(b). $B=2.5$, resonant motion.}{3}{3}

For comparison, we plot the horizontal speed and input power with $\mu_t$ increased to 100 in figures \ref{fig:SwimBvsA}(c) and (d). The curves are much smoother as noted previously. As $A$ increases from 0.03 to 0.1, the speed increases almost in proportion to $A$. This is consistent with the traveling wave solutions we have found, since the traveling wave amplitude and wavelength are both proportional to $A$. Since the foil speed is approximately the wavelength divided by the period (fixed to 1), it too is proportional to $A$. As $B\rightarrow\infty$, where the foil becomes closer to a rigid plate, the horizontal speed eventually goes to zero as discussed previously. At larger $A$ (0.3 and 0.5), the stronger nonlinearities in the equations lead to nonlinear changes in $\bar{u}$ (figure \ref{fig:SwimBvsA}(c)) and $\bar{P}$ (figure \ref{fig:SwimBvsA}(d)).

\section{Conclusion}
In this work we have studied the dynamics of an elastic foil in a frictional environment, heaved sinusoidally in time at the leading edge. The system is a model for locomotion in a frictional environment. To understand the basic physics, we began with the case where the base does not locomote horizontally. The foil dynamics depend on three key parameters: the heaving amplitude $A$, the bending rigidity $B$, and the frictional coefficients (set to the same constant $\mu$ for simplicity). With $\mu = 0$ and small $A$, we obtain the well-known case of an elastic beam in a vacuum, whose motion is a transient term, a superposition of eigenmodes at the natural frequencies set by the initial condition, together with a term set by the applied heaving motion. With nonzero $\mu$ and small $A$, the transient is damped, leaving a periodic solution. At larger $A$, there is a rich set of nonlinear behaviors that can be summarized succinctly in phase space. As $B$ is decreased from large to small, the foil transitions from a motion periodic with the driving period to a non-periodic, chaotic motion. For a band of $\mu$ values near unity, the transition passes through a set of states that are periodic at various multiples of the heaving period, and with or without bilateral symmetry. As $A$ increases from small to large, the same types of transitions occur, again with $N$-periodic states appearing at a finite band of $\mu$ values. At zero $\mu$ and small $A$, resonances occur at particular $B$ values. The resonant peaks are damped nonlinearly with increasing $\mu$ and $A$, and bistability is observed at large $A$.

Allowing the base to translate freely in $x$, and allowing different tangential and transverse friction coefficients ($\mu_f \neq \mu_t$), we find that many of the same dynamics occur (e.g. chaotic motions, resonances) together with small horizontal velocities ($\lesssim 0.1$ body lengths per period) if the $\mu_f$ and $\mu_t$ do not differ much in magnitude. In the regime $\mu_t > 1 \gg \mu_f$ corresponding to wheeled snake robots, the foil spontaneously adopts a traveling wave motion with high speed $\sim \mu_t^{1/4}$, though the input power grows faster, $\sim \mu_t^{5/12}$. We find that the motion has a boundary layer form near the leading edge in powers of $\mu_t$, consistent with the speed and input power scalings. The input power scaling in particular depends on the clamped boundary condition, which sets the slope to zero at the leading edge. We hypothesize that other leading edge conditions, such as a pinned leading edge (zero curvature), may lead to different scalings and perhaps higher locomotor efficiency. We leave this for future study, as well as comparisons with experimental work, and inclusion of proprioceptive feedback from the environment into the driving motion, as employed in other locomotion studies \cite{gazzola2015gait,tytell2010interactions,pehlevan2016integrative}.

We also point out that foil motions found here are remarkably similar to those which were found to be optimal for locomotor efficiency when the foil motion is fully prescribed at large transverse friction \cite{alben2013optimizing,wang2014optimizing}. Both motions are approximate traveling waves, and in both cases the foil slope $\partial_x y \sim \mu_t^{-1/4}$. In the present work, the input power grows rapidly with $\mu_t$, $\sim \mu_t^{5/12}$, due mainly to the necessary deviation from a traveling wave at the leading edge. For the optimal motion, the leading edge amplitude decays as $\mu_t^{-1/4}$ (versus $O(1)$ here), and the input power has the same $\mu_t$-scaling as the forward speed (both are $O(1)$). For more advantageous choices of leading edge heaving and pitching, the passive elastic foil could approach the optimal foil's performance with respect to $\mu_t$.

\begin{acknowledgments}
We thank Angelia Wang for performing simulations in a preliminary study of the fixed body dynamics. This work was supported by the National Science Foundation.
\end{acknowledgments}

\appendix
\section{Numerical solution at the first time step}
We use a second-order (BDF) discretization for the time-derivative in the numerical method. Since the discretization requires two previous time step solutions, we need to adjust the method at the first step. We give $\kappa_0$ and
$\partial_t\kappa_0$ as initial conditions. We obtain $\dtt\kappa_0$ using the central difference scheme with a guess on $\kappa_1$, and obtain $\partial_t\zeta_0$ and $\dtt\zeta_0$
by integrals. The Broyden's method is then applied at step $0$ to get the correct $\kappa_1$ and $\zeta_1$. The regular procedures can be continued after the first step to obtain
further $\kappa_n$ and $\zeta_n$.
\section{Zero Friction Model}
When the heaving amplitude $A$ is small, the nonlinear foil equation can be linearized as
\bq
\dtt y(x,t)=-B\partial_x^4y \label{eq:linbeama},
\eq
with friction coefficients set to zero. The boundary conditions become:
\bq
y(0,t)=A\sin(2\pi t),\ \partial_xy(0,t)=0;\quad \partial_x^2y(1,t)=\partial_x^3y(1,t)=0
\eq
and the initial conditions are:
\bq
y(x,0)=y_0(x),\quad \dt y(x,0)=\dt y_0(x)\label{eq:initconda}
\eq
This equation can be solved analytically by using separation of variables.

We first rewrite the solution in the form $y(x,t)=u(x,t)+v(x,t)$, where
$v(x,t)=A\sin(2\pi t)$. Then $u(x,t)$ satisfies a nonhomogeneous equation with homogeneous boundary conditions:
\bq
\dtt u(x,t)=-B\partial_x^4u+4\pi^2A\sin(2\pi t)\label{eq:linbeam}
\eq
The solution of equation (\ref{eq:linbeam}) can be represented as a series of eigenfunctions:
\bq
u(x,t)=\sum\limits_{i=1}^\infty A_i(t)\phi_i(x)
\eq
The eigenfunctions $\phi_i(x)$ correspond to the modes of a cantilevered beam \cite{erturk2008mechanical}:
\bq
\phi_i(x)=\cosh(\omega_ix)-\cos(\omega_ix)+\displaystyle\frac{\cosh\omega_i+\cos\omega_i}{\sinh\omega_i+\sin\omega_i}\left(\sin(\omega_ix)-\sinh(\omega_ix)\right)
\eq
and the eigenvalues are the roots of the nonlinear equation:
\bq
\cosh(\omega_i)\cos(\omega_i)+1=0
\eq
The eigenfunctions are orthogonal, i.e.,
\bq
\int_0^1\phi_i(x)\phi_j(x)dx=0, \quad i\neq j.
\eq
The time-dependent coefficients $A_i(t)$ therefore satisfy the nonhomogeneous ODEs:
\bq
\displaystyle\frac{d^2A_i(t)}{dt^2}+\lambda_iA_i(t)=\frac{\int_0^1\phi_i(x)dx}{\int_0^1\phi_i^2(x)dx}4\pi^2A\sin(2\pi t)\label{eq:odeai}
\eq
where $\lambda_i=B\omega_i^4$, which is also related to the eigenvalues of the vibration system. The solution of the ODE is in the form:
\bq
A_i(t)=B_i\sin(2\pi t)+C_i\cos(\sqrt{\lambda_i}t)+D_i\sin(\sqrt{\lambda_i}t)
\eq
By applying equation (\ref{eq:odeai}) and the initial conditions (\ref{eq:initconda}), we obtain the following coefficients:
\bqa
B_i=\displaystyle\frac{4\pi^2A\int_0^1\phi_i(x)dx}{(-4\pi^2+\lambda_i)\int_0^1\phi_i^2(x)dx}\\
C_i=\displaystyle\frac{\int_0^1y_0(x)\phi_i(x)dx}{\int_0^1\phi_i^2(x)dx}\\
D_i=\displaystyle\frac{1}{\sqrt{\lambda_i}}\left(\frac{\int_0^1(\dt y_0(x)-2\pi A)\phi_i(x)dx}{\int_0^1\phi_i^2(x)dx}-2\pi B_i\right)
\eqa
when $\lambda_i=B\omega_i^4\neq 4\pi^2$ (not in the resonant peaks). Therefore, the deflection of the linearized model is given by
\bq
y(x,t)=\sum\limits_{i=1}^\infty A_i(t)\phi_i(x)+A\sin(2\pi t)
\eq
The linearized model approximates
foil deflection well when the amplitude $A$ is small. In figure \ref{fig:lincomp} (a), we choose the initial conditions as $\displaystyle y_0(x)=0.01(-\frac{1}{24}x^4+\frac{1}{6}x^3-\frac{1}{4}x^2)$
and $\displaystyle\dt y_0(x)=0.01(-\frac{1}{24}x^4+\frac{1}{6}x^3-\frac{1}{4}x^2)+2\pi A$ and compare the the trailing edge displacement (free end) computed by the
numerical simulation and the linearized model until $t=20$. The other parameters used here are $A=0.01$ and $B=1$. In figure \ref{fig:lincomp}(b), we plot the spectrum of the frequency based on the free end displacement.
\Etwofigs{lmcomp1}{lmfreq}{\label{fig:lincomp}(a). Free end displacement $y$ vs. time $t$. The solid line denotes the numerical simulation result, and the dashed line denotes
the analytical result for the linearized model for $A=0.01$ and $B=1$. (b) Corresponding spectrum amplitude $|\hat{f}|$ vs. frequency.}{3}{3}
The simulation and the analytical results agree well. The coefficients $A_i$ converges to zero quickly, and the first natural vibration mode dominates as shown
in the frequency spectrum plot.

Since the coefficients $C_i$ and $D_i$ depend on the initial conditions, we can choose $y_0(x)$ and $\dt y_0(x)$ such that $C_i=D_i=0$. For example,
$y_0(x)=0$, and $\dt y_0(x)=\sum\limits_{i=1}^\infty2\pi B_i\phi_i(x)+2\pi A$. Therefore, the foil deflection for the linearized model
becomes periodic in time as $y(x,t)=\sum\limits_{i=1}^\infty B_i\sin(2\pi t)\phi_i(x)+A\sin(2\pi t)$. As we increase the magnitude $A$, nonlinearity is introduced into
the system and the periodicity will be broken. In figure \ref{fig:lmcompA}, we apply the initial conditions as discussed above, and compare the spectrum
of the frequency based on the free end displacement for the linearized model and the numerical simulation. We observe another frequency spectrum which corresponds to the
first natural frequency $\omega_1$ as $A$ increases for the
numerical results.

\Eonefigs{lmcompA}{\label{fig:lmcompA}Spectrum amplitude $|\hat{f}|$ vs. frequency based on free end displacement. The solid lines denote the numerical simulation results, and the dashed lines denote
the analytical linearized model for $B=1$ and various $A=0.01,0.05$ and 0.1.}{3}

\end{document}

%% file: macros.tex

\newcommand{\bq}{\begin{equation}}
\newcommand{\eq}{\end{equation}}
\newcommand{\bqs}{\begin{equation*}}
\newcommand{\eqs}{\end{equation*}}
\newcommand{\bqa}{\begin{eqnarray}}
\newcommand{\eqa}{\end{eqnarray}}
\newcommand{\bqas}{\begin{eqnarray*}}
\newcommand{\eqas}{\end{eqnarray*}}
\def\etal{{\em et al.\ }}
\newcommand{\dtt}{\partial_{tt}}
\newcommand{\ds}{\partial_{s}}
\newcommand{\dt}{\partial_t}
\newcommand{\z}{\zeta}
\newcommand{\Dfig}[2]{\includegraphics*[width=#2in]{#1.eps}}
\newcommand{\hs}{\hat{s}}
\newcommand{\hn}{\hat{n}}
\newcommand{\hy}{\hat{y}}
\newcommand{\hp}{\hat{[p]}}
\newcommand{\hv}{\hat{V}}
\newcommand{\hg}{\hat{\gamma}}
\def\a{\alpha}
\def\f{\frac}
\def\bX{{\bf X}}
\def\bF {{\bf f}}
\def\th {\theta}
\def\k {\kappa}
\def\ptZ{\partial_t\zeta}
\def\hptZ{\widehat{\partial_t\zeta}}
\newcommand{\tmu}{\tilde{\mu}}

\newenvironment{Eonefigs}[3]
{
        \begin{figure} [ht]
          \begin{center}
          \begin{tabular}{c}
              \Dfig{#1}{#3} \\
           \vspace{-.25in}
          \end{tabular}
         \caption{{\footnotesize #2}}
          \label{#1}
          \end{center}
        \vspace{-0in}
       \end{figure}
}{}
\newenvironment{Etwofigs}[5]
{
        \begin{figure} [htpb]
          \begin{center}
          \setlength{\tabcolsep}{.1in}
          \begin{tabular}{cc}
          \multicolumn{1}{l}{$\bf{(a)}$}&\multicolumn{1}{l}{$\bf{(b)}$}\\
              \Dfig{#1}{#4} & \Dfig{#2}{#5} \\
                   \vspace{-.25in}
          \end{tabular}
          \caption{\footnotesize #3}
          \label{#1}\label{#2}
          \end{center}
        \vspace{-.25in}
        \end{figure}
}{}

\newenvironment{Etwofigss}[5]
{
\begin{figure} [htpb]
          \begin{center}
          \setlength{\tabcolsep}{.1in}
          \begin{tabular}{cc}
          \multicolumn{1}{l}{$\bf{(a)}$}&\multicolumn{1}{l}{$\bf{(b)}$}\\
              \Dfig{#1}{#4} &\vspace{0.3in} \Dfig{#2}{#5} \\
                   \vspace{-.25in}
          \end{tabular}
          \caption{\footnotesize #3}
          \label{#1}\label{#2}
          \end{center}
        \vspace{-.25in}
        \end{figure}
}{}

\newenvironment{Ethreefigs}[7]
{
        \begin{figure} [htpb]
          \begin{center}
          \setlength{\tabcolsep}{.1in}
          \begin{tabular}{ccc}
          \multicolumn{1}{l}{$\bf{(a)}$}&\multicolumn{1}{l}{$\bf{(b)}$}&\multicolumn{1}{l}{$\bf{(c)}$}\\
              \Dfig{#1}{#5} & \Dfig{#2}{#6}&\Dfig{#3}{#7} \\
                    \vspace{-.25in}
          \end{tabular}
          \caption{\footnotesize #4}
          \label{#1}\label{#2}\label{#3}
          \end{center}
        \vspace{-.25in}
        \end{figure}
}{}
\newenvironment{Ethreefigsv}[7]
{
        \begin{figure} [htpb]
          \begin{center}
          \setlength{\tabcolsep}{.1in}
          \begin{tabular}{cc}
          \multicolumn{1}{l}{$\bf{(a)}$}&\multicolumn{1}{l}{$\bf{(b)}$}\\
              \Dfig{#1}{#5} & \Dfig{#2}{#6} \\
          \multicolumn{2}{l}{$\bf{(c)}$}\\
          \multicolumn{2}{c}{\Dfig{#3}{#7}}
                    \vspace{-.25in}
          \end{tabular}
          \caption{\footnotesize #4}
          \label{#1}\label{#2}\label{#3}
          \end{center}
        \vspace{-.25in}
        \end{figure}
}{}
\newenvironment{Ethreefigsvp}[7]
{
        \begin{figure} [htpb]
          \begin{center}
          \setlength{\tabcolsep}{.1in}
          \begin{tabular}{c}
          \multicolumn{1}{l}{$\bf{(a)}$}\\
          \Dfig{#1}{#5}\\
          \multicolumn{1}{l}{$\bf{(b)}$}\\
           \Dfig{#2}{#6}\\
          \multicolumn{1}{l}{$\bf{(c)}$}\\
          \Dfig{#3}{#7}
                    \vspace{-.25in}
          \end{tabular}
          \caption{\footnotesize #4}
          \label{#1}\label{#2}\label{#3}
          \end{center}
        \vspace{-.25in}
        \end{figure}
}{}
\newenvironment{Efourfigsv}[9]
{
        \begin{figure} [htpb]
          \begin{center}
           \setlength{\tabcolsep}{.1in}
          \begin{tabular}{cc}
          \multicolumn{1}{l}{$\bf{(a)}$}&\multicolumn{1}{l}{$\bf{(b)}$}\\
              \Dfig{#1}{#6} & \Dfig{#2}{#7} \\
              \multicolumn{1}{l}{$\bf{(c)}$}&\multicolumn{1}{l}{$\bf{(d)}$}\\
                     \Dfig{#3}{#8} & \Dfig{#4}{#9} \\
                   \vspace{-.25in}
          \end{tabular}
          \caption{{\footnotesize #5}}
          \label{#1}\label{#2}\label{#3}\label{#4}
          \end{center}
        \vspace{-.25in}
        \end{figure}
}{}
\newenvironment{Efourfigs}[7]
{
        \begin{figure} [htpb]
          \begin{center}
           \setlength{\tabcolsep}{.1in}
          \begin{tabular}{cc}
          \multicolumn{1}{l}{$\bf{(a)}$}&\multicolumn{1}{l}{$\bf{(b)}$}\\
              \Dfig{#1}{#6} & \Dfig{#2}{#6} \\
              \multicolumn{1}{l}{$\bf{(c)}$}&\multicolumn{1}{l}{$\bf{(d)}$}\\
                     \Dfig{#3}{#7} & \Dfig{#4}{#7} \\
                   \vspace{-.25in}
          \end{tabular}
          \caption{{\footnotesize #5}}
          \label{#1}\label{#2}\label{#3}\label{#4}
          \end{center}
        \vspace{-.25in}
        \end{figure}
}{}
\newenvironment{Efivefigs}[8]
{
        \begin{figure} [htpb]
          \begin{center}
          \setlength{\tabcolsep}{.1in}
          \begin{tabular}{ccc}
          \multicolumn{1}{l}{$\bf{(a)}$}&\multicolumn{1}{l}{$\bf{(b)}$}&\multicolumn{1}{l}{$\bf{(c)}$}\\
              \Dfig{#1}{#7} & \Dfig{#2}{#7}&\Dfig{#3}{#7}\\
            \multicolumn{1}{l}{$\bf{(d)}$}&\multicolumn{1}{l}{$\bf{(e)}$}&\\
               \Dfig{#4}{#7}& \multicolumn{2}{c}{\Dfig{#5}{#8}} \\
                    \vspace{-.25in}
          \end{tabular}
          \caption{{\footnotesize #6}}
          \label{#1}\label{#2}\label{#3}\label{#4}\label{#5}
          \end{center}
        \vspace{-.25in}
        \end{figure}
}{}

\newenvironment{Efivefigsv}[8]
{
        \begin{figure} [htpb]
          \begin{center}
          \setlength{\tabcolsep}{.1in}
          \begin{tabular}{ccc}
          \multicolumn{1}{l}{$\bf{(a)}$}&\multicolumn{1}{l}{$\bf{(b)}$}& \multicolumn{1}{l}{$\bf{(c)}$}\\
  \multirow{3}{*}[0.95in]{\Dfig{#1}{#7}} &\Dfig{#2}{#8} & \Dfig{#3}{#8}\\
&\multicolumn{1}{l}{$\bf{(d)}$}& \multicolumn{1}{l}{$\bf{(e)}$}\\
            & \Dfig{#4}{#8}&\Dfig{#5}{#8}\\
                    \vspace{-.25in}
          \end{tabular}
          \caption{{\footnotesize #6}}
          \label{#1}\label{#2}\label{#3}\label{#4}\label{#5}
          \end{center}
        \vspace{-.25in}
        \end{figure}
}{}
\newenvironment{ESixfigs}[8]
{
        \begin{figure} [htpb]
          \begin{center}
          \setlength{\tabcolsep}{.1in}
          \begin{tabular}{ccc}
          \multicolumn{1}{l}{$\bf{(a)}$}&\multicolumn{1}{l}{$\bf{(b)}$}&\multicolumn{1}{l}{$\bf{(c)}$}\\
              \Dfig{#1}{#8} & \Dfig{#2}{#8}& \Dfig{#3}{#8}\\
              \multicolumn{1}{l}{$\bf{(d)}$}&\multicolumn{1}{l}{$\bf{(e)}$}&\multicolumn{1}{l}{$\bf{(f)}$}\\
                \Dfig{#4}{#8}& \Dfig{#5}{#8}&\Dfig{#6}{#8} \\ 
                    \vspace{-.25in}
          \end{tabular}
          \caption{{\footnotesize #7}}
          \label{#1}\label{#2}\label{#3}\label{#4}\label{#5}\label{#6}
          \end{center}
        \vspace{-.25in}
        \end{figure}
}{}
\newenvironment{ESixfigsv}[8]
{
        \begin{figure} [htpb]
          \begin{center}
          \setlength{\tabcolsep}{.1in}
          \begin{tabular}{cc}
          \multicolumn{1}{l}{$\bf{(a)}$}&\multicolumn{1}{l}{$\bf{(b)}$}\\
              \Dfig{#1}{#8} & \Dfig{#2}{#8}\\
              \multicolumn{1}{l}{$\bf{(c)}$}&\multicolumn{1}{l}{$\bf{(d)}$}\\
                \Dfig{#3}{#8}& \Dfig{#4}{#8} \\ 
           \multicolumn{1}{l}{$\bf{(e)}$}&\multicolumn{1}{l}{$\bf{(f)}$}\\
              \Dfig{#5}{#8}&\Dfig{#6}{#8}
                    \vspace{-.25in}
          \end{tabular}
          \caption{{\footnotesize #7}}
          \label{#1}\label{#2}\label{#3}\label{#4}\label{#5}\label{#6}
          \end{center}
        \vspace{-.25in}
        \end{figure}
}{}
\def\Xint#1{\mathchoice
{\XXint\displaystyle\textstyle{#1}}%
{\XXint\textstyle\scriptstyle{#1}}%
{\XXint\scriptstyle\scriptscriptstyle{#1}}%
{\XXint\scriptscriptstyle\scriptscriptstyle{#1}}%
\!\int}
\def\XXint#1#2#3{{\setbox0=\hbox{$#1{#2#3}{\int}$}
\vcenter{\hbox{$#2#3$}}\kern-.5\wd0}}
\def\ddashint{\Xint=}
\def\dashint{\Xint-}